\documentclass[ reprint, superscriptaddress, amsmath,amssymb, aps]{revtex4-1}
\usepackage{graphicx}% Include figure files
\usepackage{dcolumn}% Align table columns on decimal point
\usepackage{amsthm}
\usepackage{bm}% bold math

\pdfoutput=1
\usepackage{graphicx,graphics,epsfig,subfigure,times,bm,bbm,amssymb,amsmath,amsthm,mathrsfs,MnSymbol}
\usepackage{gensymb}
\usepackage{amsfonts}
\usepackage[matrix,frame,arrow]{xypic}
\usepackage[pdfstartview=FitH]{hyperref}
\hypersetup{
    colorlinks=true,       % false: boxed links; true: colored links
    linkcolor=red,          % color of internal links
    citecolor=magenta,        % color of links to bibliography
    filecolor=magenta,      % color of file links
    urlcolor=cyan,           % color of external links
    runcolor=cyan}
\usepackage{epstopdf}
\usepackage[pdftex]{color}
\usepackage{braket}%Dirac Notation in QM
\usepackage{enumerate}
\usepackage[normalem]{ulem}
\usepackage[usenames,dvipsnames]{xcolor}
\usepackage{multirow}
\usepackage{mathtools}

\definecolor{orange}{rgb}{1,0.5,0}

\newcommand{\ignore}[1]{}
\usepackage{geometry}

\ignore{
\documentclass[eprintnumbers,amsmath,amssymb,onecolumn,a4paper,caption ]{article}
\usepackage{amsfonts}
\usepackage{amssymb}
\usepackage{mathrsfs}
\usepackage{mathbbold}
\usepackage{bbm}
\usepackage{mathrsfs}
\usepackage{dcolumn}% Align table columns on decimal point
\usepackage{bm}% bold math
\usepackage{times,epsfig,amssymb,amsmath}
\usepackage{float}
\usepackage{subfigure}
\usepackage{geometry}\geometry{left=2.5cm,right=2.5cm,top=3cm,bottom=3cm}

\usepackage{color}

}

\begin{document}

\title{Quench dynamics of entanglement spectra and topological superconducting phases in a long-range Hamiltonian}

\author{Kaixiang Su}
\affiliation{School of Physics, Peking University, Beijing 100871, China}

\author{Zheng-Hang Sun}
\email{zhsun@iphy.ac.cn}
\affiliation{Institute of Physics, Chinese Academy of Sciences, Beijing 100190, China}
\affiliation{School of Physical Sciences, University of Chinese Academy of Sciences, Beijing 100190, China}

\author{Heng Fan}
\email{hfan@iphy.ac.cn}
\affiliation{Institute of Physics, Chinese Academy of Sciences, Beijing 100190, China}
\affiliation{School of Physical Sciences, University of Chinese Academy of Sciences, Beijing 100190, China}
\affiliation{CAS Central of Excellence in Topological Quantum Computation, Beijing 100190, China}

\begin{abstract}
We study the quench dynamics of entanglement spectra in the Kitaev chain with variable-range pairing quantified by power-law decay rate $\alpha$. Considering the post-quench Hamiltonians with flat bands, we demonstrate that the presence of entanglement-spectrum crossings during its dynamics is able to characterize the topological phase transitions (TPTs) in both short-range ($\alpha>1$) and long-range ($\alpha<1$) sectors. Novel properties of entanglement-spectrum dynamics are revealed for the quench protocols in the long-range sector or with $\alpha$ as the quench parameter. In particular, when the lowest upper-half entanglement-spectrum value of initial Hamiltonian is smaller than the final one, the TPTs can also be diagnosed by the difference between the lowest two upper-half entanglement-spectrum values if the half-way winding number is not equal to that of the initial Hamiltonian. Moreover, we discuss the stability of characterizing the TPTs via entanglement-spectrum crossings against energy dispersion in the long-range model.
\end{abstract}
\pacs{Valid PACS appear here}
\maketitle

\section{Introduction}
Topological superconductors have attracted considerable interests in recent years. The Majorana zero modes in those systems, being robust against disorder~\cite{rev1,rev2,exp1,exp2}, play a key role in the realization of topological quantum computation~\cite{qtc1,qtc2,qtc3,qtc4,qtc5}. One of the most intriguing topological superconductors is the Kitaev chain with long-range $p$-wave pairing terms, where novel topological phases with fractional winding numbers and  massive Dirac edge states are found~\cite{long_range}. More importantly, the Hamiltonian can be realized in magnetic atomic chains~\cite{exp1,exp3,exp4,exp5} with the long-range pairing induced by magnetic impurities~\cite{mag1,mag2,mag3,mag4,mag5,mag6,mag7}.

The characterization of topological phase transitions (TPTs), beyond the Laudau symmetry-breaking paradigm, is also of great significance. From the perspective of quantum information, it has been shown that the quantum coherence~\cite{coherence,coherence_add}, multipartite entanglement~\cite{QFI1,QFI2,QFI3,QFI4}, entanglement entropy~\cite{EE1,EE2,EE3,EE_add} and entanglement spectrum~\cite{ES1,ES2,ES3} of the ground state can detect the TPTs.
Recently, the rapid developments of quantum simulation based on ultracold atoms~\cite{ua_sum,ua1,ua2}, trapped ions~\cite{ti_sum} and superconducting qubits~\cite{sq1,sq2} have stimulated the study of quench dynamics. It is therefore natural to extend the characterization of TPTs to an out-of-equilibrium regime~\cite{noneq1,noneq2,noneq3,noneq4,noneq5}.
For instance, the quench dynamics of entanglement spectra in the topological insulators and superconductors with nearest-neighbor terms, involving the standard Su-Schrieffer-Heeger model~\cite{SSH} and Kitaev chain~\cite{Kitaevchain} are studied, suggesting its close relationship with TPTs~\cite{ESnoneq1,ESnoneq2,ESnoneq3}.
Nevertheless, the investigation of the entanglement-spectrum non-equilibrium behaviors in long-range models remains limited, and the methods useful in short-range systems can be further generalized.

In this work, we explore the quench dynamics of entanglement spectra in the Kitaev chain with long-range pairing, whose topological phase diagram is more complex than the previously studied models~\cite{long_range,EE3}. The values of entanglement spectra can be efficiently measured via the quantum state tomography implemented in various artificially-engineered platforms~\cite{qst1,qst2,qst3}. Therefore, our results can be tested by state-of-art quantum simulation experiments. The remainder is organized as follows. In Sec. \uppercase\expandafter{\romannumeral2}, we briefly review the Hamiltonian and the topological phases of this model, and the definition of the single-particle entanglement spectrum. We also calculate the lowest upper-half entanglement-spectrum value of ground state in this model and present a physical picture of our work. In the following, the lowest one refers to the \emph{lowest upper-half one}, the meaning of which will be explained in Sec. \uppercase\expandafter{\romannumeral2}. B. In Sec. \uppercase\expandafter{\romannumeral3}, we study the quench dynamics of entanglement spectra in the Kitaev chain with variable range pairing, revealing several novel nonequilibrium properties of entanglement spectra and demonstrating that the TPTs in the long-range  Hamiltonian can be characterized by the quench dynamics of entanglement spectra. In Sec. \uppercase\expandafter{\romannumeral4}, we conclude and provide some outlooks.

\section{Preliminary}
\subsection{The model}
We focus on the long-range Kitaev chain with power-law decay pairing terms~\cite{long_range,EE3}, as a generalization of the standard Kitaev chain with only nearest-neighbor terms~\cite{Kitaevchain}. The Hamiltonian reads
\begin{eqnarray}
H = &-&\frac{t}{2}\sum_{i=1}^{N}(c_{i}^{\dagger}c_{i+1}+\text{H.c.}) - \mu\sum_{i=1}^{N}(c_{i}^{\dagger}c_{i}-\frac{1}{2}) \nonumber \\
&+& \frac{\Delta}{2}\sum_{i=1}^{N}\sum_{l=1}^{N-i}\frac{1}{d_{l}^{\alpha}}(c_{i}c_{i+l} + \text{H.c.})
,
\label{Hamiltonian}
\end{eqnarray}
where $c_{i}$ ($c_{i}^{\dagger}$) denotes the annihilation (creation) fermion operator at each site $i$, $N$ is the length of Kitaev chain, and $t$ and $\mu$ represent the hopping amplitude and the chemical potential, respectively. The amplitude of pairing $\Delta$ decay with the parameter $\alpha$ (the decay rate) of the distance $d_{l}$. Here, the antiperiodic boundary condition $c_{i+N}=-c_{i}$ (see Appendix A for the reason behind the choice of the antiperiodic boundary condition), and the condition of closed chain, i.e., $d_{l}=l$ for $l\in[1,N/2]$, while $d_{l}=N-l$ for $l\in[N/2,N]$, are adopted.

\begin{figure}
	\centering
	\includegraphics[width=1\linewidth]{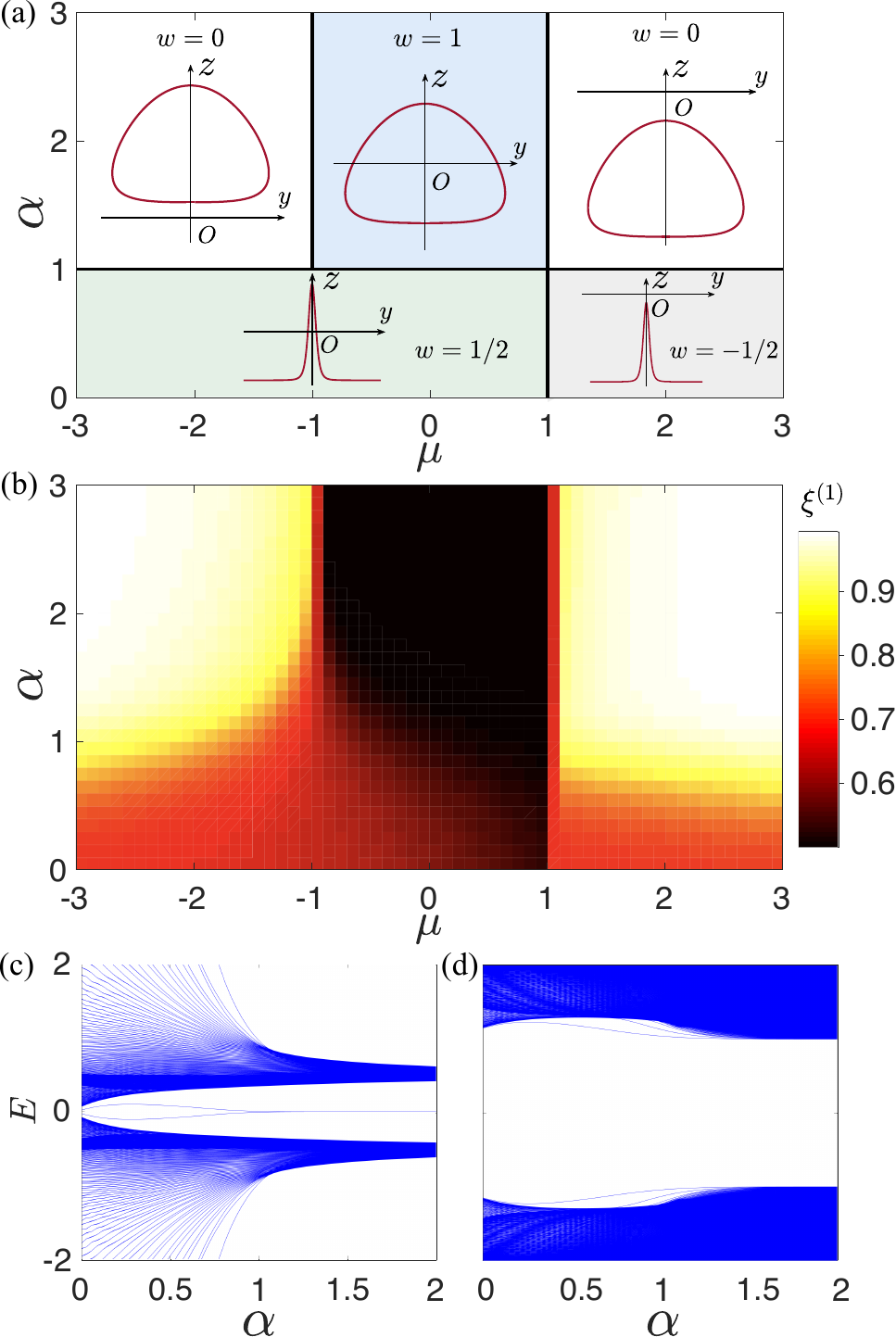}\\
	\caption{(a) Phase diagram of the Hamiltonian (\ref{Hamiltonian}) with $t=-\Delta=1$, characterized by the winding number and the trajectory of $\bm{d}_{k}$. (b) The lowest entanglement-spectrum value $\xi^{(1)}$ for the ground state of Hamiltonian (\ref{Hamiltonian}) with $t=-\Delta=1$ as a function of $\alpha$ and $\mu$. (c) The energy spectra for the Hamiltonian (\ref{Hamiltonian}) with $t=-\Delta=1$ and $\mu=0$ as a function of $\alpha$. (d) is similar to (c) but with $\mu=-3$. Here, the system size is $N=200$. We emphasize that the depicted energy spectra are obtained by directly diagonalizing the Hamiltonian with open boundary conditions.}\label{Fig1}
\end{figure}

By switching to the momentum space via the Fourier transformation,
the Hamiltonian (\ref{Hamiltonian}) can be written as $H=\sum_{k}\epsilon_{k}\Psi_{k}^{\dagger}(\bm{d}_{k}\cdot \bm{\sigma})\Psi_{k}$ with $\Psi_{k}^{\dagger}$ as the Nambu spinor and $\bm{\sigma}$ as the Pauli vector. The winding vector is
\begin{eqnarray}
\bm{d}_{k} &=& (d_{k}^{x},d_{k}^{y},d_{k}^{z}) \nonumber \\
&=& (0,-\frac{1}{2}\Delta f_{\alpha}(k),-\frac{\mu+t\cos k}{2})
\label{Anderson}
\end{eqnarray}
where $k=(2\pi/N)(n+1/2)$ $(n=0,1,...,N-1)$ and $f_{\alpha}(k) = \sum_{i=1}^{N-1} \sin (kl)/ d_l^\alpha$. The energy spectra is then $\epsilon_{k} = |\bm{d}_{k}|$.
The topological phases in this model can be characterized by the $\textbf{Z}$ topological invariant winding number~\cite{winding} defined as
\begin{eqnarray}
w = \frac{1}{2\pi} \oint dk \left( \frac{\partial_{k} d_{k}^{z}}{d_{k}^{y}} \right),
\end{eqnarray}
which can be rewritten as $w=(1/2\pi)\oint (ydz-zdy )/|\bm{d}_{k}|^{2}$ with $y$ ($z$) as the $y$ $(z)-$component of Eq. (\ref{Anderson}). Intuitively, it counts how many times $\bm{d}_{k}$ loops around the origin in the $y-z$ plane. Thus,
the winding number can also be obtained by simply plotting the trajectory of the winding vector Eq. (\ref{Anderson}). The phase diagram of the Hamiltonian (\ref{Hamiltonian}) with $t=-\Delta=1$ (which is fixed in the rest of this work) is shown in Fig. \ref{Fig1}(a). The topological phases are similar to the Kitaev chain with the nearest pairing term in the short-range sector ($\alpha>1$), while different from the conventional Kitaev chain in the long-range sector ($\alpha<1$). In particular, the topological phase with a massive Dirac edge mode has winding number $1/2$, however the winding number of the trivial phase is $-1/2$.

Our quench protocol is as follows: The initial state is prepared as the ground state of a Hamiltonian $H^{i}$ (the initial Hamiltonian). We then evolve it with the final Hamiltonian $H^{f}$.  Notice that the evolved state can be viewed as the ground state of the following Hamiltonian
\begin{eqnarray}
H(t) = e^{-iH^ft}H^ie^{iH^ft}
\label{dynamic0}
\end{eqnarray}
In the momentum space, this Hamiltonian can be similarly represented by its winding vector, whose dynamics reads~\cite{ESnoneq1}
\begin{eqnarray}
\partial_{t} \bm{d}_{k}(t) = 2 \bm{d}^{f}_{k} \times \bm{d}_{k}(t),
\label{dynamic1}
\end{eqnarray}
which can be explicitly solved as
\begin{eqnarray}
\bm{d}_{k}(t) = &[&1-\cos (2|\bm{d}_{k}^{f}|t)][\bm{d}^{i}_{k} \cdot \bm{n}_{k}^{f}]\bm{n}_{k}^{f} \nonumber \\
&+& \cos (2|\bm{d}_{k}^{f}|t)\bm{d}_{k}^{i} +\sin (2|\bm{d}_{k}^{f}|t)[\bm{d}_{k}^{i} \times \bm{n}_{k}^{f}]
\label{dynamic2}
\end{eqnarray}
with $\bm{d}_{k}^{i}$ ($\bm{d}_{k}^{f}$) referring to the winding vector of $H^{i}$ ($H^{f}$), and $\bm{n}_{k}^{f} \equiv -\bm{d}_{k}^{f}/|\bm{d}_{k}^{f}| $.

Here, we briefly mention the concept of band flattening. A flat-band Hamiltonian is obtained by a continuous deformation of the original Hamiltonian, setting the eigenvalues as $\pm 1$. The winding vectors of the flat-band Hamiltonian are all of length 1.

\subsection{Entanglement spectrum}
Next, we present the definition of entanglement spectrum. In a fermionic model, the reduced density matrix of a subsystem $\mathcal{A}$ has the form ~\cite{EE_Hamiltonian}
\begin{eqnarray}
\rho_{\mathcal{A}} \propto \exp(-\sum_{q}\Omega_{q} \gamma_{q}^{\dagger}\gamma_{q}),
\label{EDM}
\end{eqnarray}
which is closely related to the single-particle entanglement spectrum defined as~\cite{Es_add1}
\begin{eqnarray}
\xi_{q} \equiv 1/[1+\exp(-\Omega_{q})]
\label{ES_def}
\end{eqnarray}
with $\Omega_{q}>0$. The method of calculating the $\Omega_{q}$ for the system Eq. (\ref{Hamiltonian}) and its dynamics Eq. (\ref{dynamic0}) is presented in Appendix B. In this work, we mainly discuss the first and second lowest entanglement-spectrum value denoted as $\xi^{(1)}$ and $\xi^{(2)}$ respectively.

An exact correspondence between the entanglement spectrum and the spectrum of physical edge modes is given in Ref.~\cite{ES3}. Specifically, entanglement spectrum can be casted into the form
\begin{eqnarray}
\xi_q = \frac{1}{2}+\frac{\lambda_q}{2}.
\label{ES_energy}
\end{eqnarray}
In the following, we focus on the single-particle entanglement spectra whose values are larger than $1/2$, i.e., the upper-half ones, since all the spectra come in pair and are related by a particle-hole transformation.
The $\lambda_q$ equals the energy spectrum of the corresponding spectrally flattened Hamiltonian restricted to subsystem $\mathcal{A}$. Since band-flattening in general does not change the topology of the system, topological edge modes can be directly read off by looking at the low-lying entanglement spectra. As an example, topological superconductors in BDI class is characterized by a \textbf{Z} topological index~\cite{Kitaev_add}. This topological invariant is directly related to the winding number $w$, and gives the number of massless edge modes on one edge ($\lambda_{q}=0$). In this case, the $\xi^{(1)}$ will be $1/2$, and we refer to this phenomena as the entanglement-spectrum crossing(s).

Before we study the quench dynamics of entanglement spectra, we first illustrate the properties of $\xi^{(1)}$ for the ground states as a benchmark.
As shown in Fig. \ref{Fig1}(b), in the short-range sector ($\alpha>1$), the topological phase can be characterized by the $\xi^{(1)}$ for the ground states. The $\xi^{(1)} \simeq 0.5$ for the topological phase while $\xi^{(1)} \simeq 1$ in the trivial phase, corresponding to the presence or absence of massless edge modes.
This difference is however less prominent in the long-range sector ($\alpha<1$). Since the long range topological phase features a massive edge mode, the $\xi^{(1)}$ in general is not close to 0.5. Nevertheless, we could still pinpoint the phase boundary $\mu_{c}=1$ by the sharp change of $\xi^{(1)}$.

We also plot the energy spectra as a function of $\alpha$ for $\mu=0$ and $\mu=-3$ in Fig.~\ref{Fig1}(c) and (d) respectively. With $\mu=0$, we observe massive Dirac fermions when $\alpha<1$ and one massless edge state when $\alpha>1$. However, with $\mu=-3$, there is no massless edge state when $\alpha>1$. The results of energy spectrum reveal the mechanism of the TPTs driven by $\alpha$. It can be recognized that at the critical point $\alpha_{c}=1$, there is a gap of the energy spectrum, which can explain that the phase boundary $\alpha_{c}=1$ is less distinguishable, and the change of $\xi^{(1)}$ is continuous when crossing the phase boundary (shown in Fig.~\ref{Fig1}(b)). In addition, the more obvious phase boundary $\mu_{c}=1$ in the long-range sector also corresponds to the behaviors of energy spectrum. In Ref.~\cite{long_range}, it is seen that the energy spectrum of the system becomes gapless at the critical point $\mu_{c}=1$ for both short and long-range sector.

\subsection{Physical picture}
To focus on the topological properties of the quench dynamics, we can restrict ourselves to the case where $H^f$ is a band-flattened Hamiltonian. For concreteness, we can take the length of winding vector $\epsilon_{k} = |\bm{d}_{k}|$ to be 1. With this condition the winding vectors $d_k(t)$ will process at the same velocity. The Hamiltonian $H(t)$ is thus time-periodic. It has been shown that for short-range systems, entanglement-spectrum crossings will appear half-way through the time evolution, i.e., $t=\pi/2$, and the number of crossings are related to the dynamical topological indices. Here, we present a physical picture to relate the number of crossings to the topological indices of $H^i$ and $H^f$. Utilizing this picture, we will then present and analyze most of our results on the long-range Kitaev chains and explain how the entanglement-spectrum behaviors differ from the short-range case.

The topological superconductor considered here belongs to the BDI class, and the winding vectors will lie on the $y-z$ plane due to the symmetry constraints. In our quench protocol, $\bm{d}^i$ and $\bm{d}^f$ will satisfy this condition, while $\bm{d}(t)$ will not lie in one plane for an arbitrary time instant. The only exception will be $t=\pi/2$ and $t=0 (\pi)$, while the latter case is simply $\bm{d}^i$ itself. The half-way entanglement-spectrum crossings is then associated with the half-way winding vector, i.e., $\bm{d}(\pi/2)$. Since the $\bm{d}(\pi/2)$ still possesses the symmetry constraints, the number of its edge modes can be directly counted by the winding number. We can then view the half-way entanglement spectra as characterizing the topology of this half-way Hamiltonian.

\section{Results}
\subsection{Chemical potential $\mu$ as the quench parameter}
We first focus on the entanglement-spectrum dynamical properties with the quench protocols where the chemical potential $\mu$ is chosen as the quench parameter, i.e., $\mu=\mu_{i}\rightarrow \mu_{f}$, and other parameters are fixed. The $\mu_{i}$ and $\mu_{f}$ refer to the chemical potential of the initial and final Hamiltonian respectively.

\begin{figure}
	\centering
	\includegraphics[width=1\linewidth]{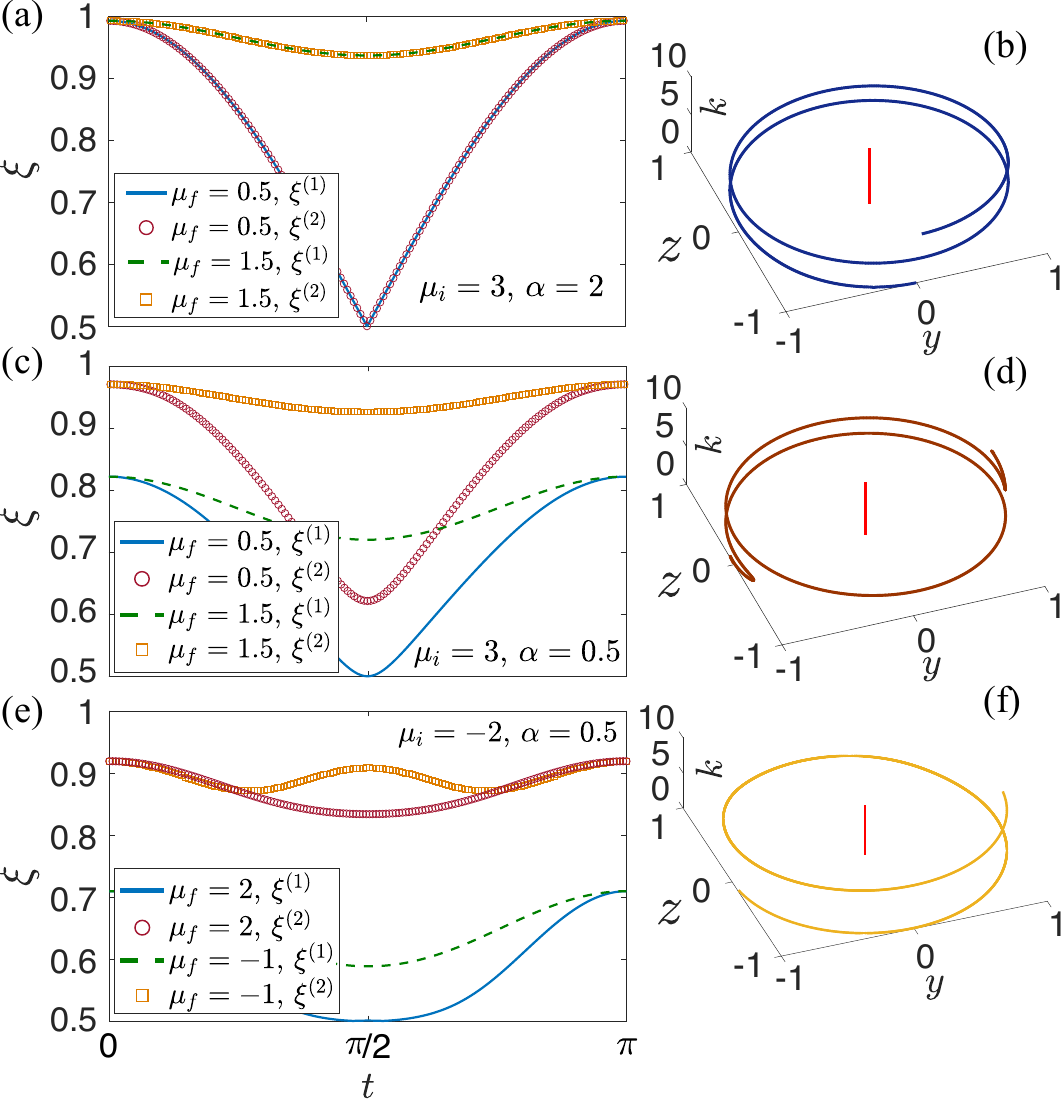}\\
	\caption{(a) Time evolution of the first and second lowest entanglement-spectrum value $\xi^{(1)}$ and $\xi^{(2)}$ with the quench protocol $\mu=3\rightarrow 1.5$ and $0.5$ in the Hamiltonian (\ref{Hamiltonian}) with $\alpha=2$ and $\Delta=t=1$. (b) The trajectory of $\bm{d}_{k}$ at time $t=\pi/2$ for the quench protocol in (a) with $\mu_{f}=1.5$. (c) and (d) are similar to (a) and (b), respectively, but with another quench protocol $\mu=3\rightarrow 1.5$ and $0.5$ in the Hamiltonian (\ref{Hamiltonian}) with $\alpha=0.5$  (e) is similar to (a), but with another quench protocol $\mu=-2\rightarrow -1$ and $2$ in the Hamiltonian (\ref{Hamiltonian}) with $\alpha=0.5$. (f) is similar to (b) but for the quench protocol in (e) with $\mu_{f}=2$.}\label{Fig2}
\end{figure}

\begin{figure*}
	\centering
	\includegraphics[width=1\linewidth]{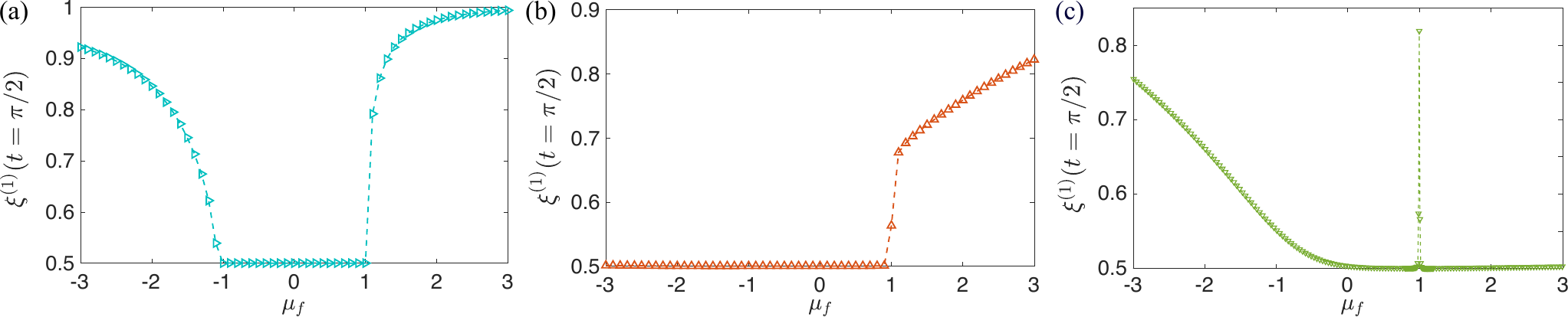}\\
	\caption{(a) The lowest entanglement-spectrum value for the quench state at $t=\pi/2$, i.e., $\xi^{(1)}(t=\pi/2)$, with the quench protocol $\mu=3\rightarrow \mu_{f}$ as a function of $\mu_{f}$ in the Hamiltonian (\ref{Hamiltonian}) with $\alpha=2$. Here, the critical points are $\mu_{c}=\pm 1$. (b) is similar to (a) but in the Hamiltonian (\ref{Hamiltonian}) with $\alpha=0.5$, and the critical point is $\mu_{c}=1$ (c) is similar to (a) but with the quench protocol $\mu=-3\rightarrow \mu_{f}$ in the Hamiltonian (\ref{Hamiltonian}) with $\alpha=0.5$, and the critical point is $\mu_{c}=1$.}\label{Fig3}%\end{figure*}
\end{figure*}

As a warm up, we review the results in the systems with nearest-neighbor interactions. Taking $H^i$ to be in the trivial phase, there will always be two degenerate entanglement-spectrum crossings as long as $H^f$ belongs to the topological phase. If $H^f$ is also in the trivial phase, even the lowest entanglement-spectrum value is far away from 1/2. Thus the entanglement-spectrum crossings provide a distinctive signature for diagnosing topological phases~\cite{ESnoneq1,ESnoneq2,ESnoneq3}. We now show that similar entanglement-spectrum behaviors can also be observed in our Hamitonian (\ref{Hamiltonian}) in the short-range sector ($\alpha>1$). We study the quench dynamics with the parameters in (\ref{Hamiltonian}) as $\Delta=t=1$, $\alpha=2$ and $\mu_{i}=3$. The results are depicted in Fig. \ref{Fig2}(a). It can be seen that the $\xi^{(1)}$ approaches $0.5$ for $\mu_{f}=1.5$ while it remains a larger value ($>0.9$) for $\mu_{f}=0.5$, characterizing the presence or absence of a TPT. Moreover, the degeneracy of two lowest entanglement-spectrum values, i.e., $\xi^{(1)}=\xi^{(2)}$, is observed for both $\mu_{f}=1.5$ and $0.5$, which can be interpreted by looking at the half-way winding vector. In Fig. \ref{Fig2}(b), we plot the trajectory of $\bm{d}_{k}$ at the time $t=\pi/2$, indicating that the winding number of the quenched state is $w=2$, thus explaining the doubly degenerate crossings. It has also been demonstrated that the quench dynamics of entanglement spectra with a nontrivial $H^i$, such as one with $w=1$, can not reveal the signatures of TPTs~\cite{ESnoneq1}, which is still hold for the short-range sector of the Hamiltonian Eq. (\ref{Hamiltonian}) (see Appendix C).

Furthermore, we study the entanglement spectra in the long-range system with $\alpha=0.5$ and $\mu_{i}=3$, and novel entanglement-spectrum behaviors are observed. As shown in \ref{Fig2}(c), the entanglement-spectrum crossing is observed when quenching across the critical line $\mu_c = 1$ ($\mu_{f}$=0.5), and is absent when staying in the same phase ($\mu_f = 1.5$). However, different from the results in Fig. \ref{Fig2}(a), the degeneracy of $\xi^{(1)}$ and $\xi^{(2)}$ is destroyed. This can be traced to the fact that the half-way winding vector is different from that of the short-range case. In the long-range sector, the half-way winding number is $-3/2$ (see Fig. \ref{Fig2}(d)), and there is presumably only one pair of massless Majorana modes, corresponding to the non-degenerate entanglement spectra. To further understand the topological properties of the phase with winding number $w=-3/2$, in Appendix D, we construct an extended Kitaev chain with long-range pairing where the topological phase with $w=\pm 3/2$ exists, ensuring that there is one pair of massless edge modes in the topological phase with $w=\pm 3/2$.

Conventionally, the initial state is chosen as a topologically trivial ground state~\cite{ESnoneq1,ESnoneq2,ESnoneq3}. To investigate if the above statements still hold in the long-range sector, we choose our $H^i$ to be in the topologically nontrivial phase with massive Dirac edge states and winding number $w=1/2$, while our $H^f$ is a trivial one with $w=-1/2$. Specifically, we quench the chemical potential $\mu=-2\rightarrow -1$ and $2$ separately with $\alpha=0.5$. It is quite remarkable that the TPT can still be characterized by the dynamics of entanglement spectra in this case, as shown in Fig. \ref{Fig2}(e). Actually, the half-way winding number of the quench in Fig. \ref{Fig2}(e) is $3/2$ (see Fig. \ref{Fig2}(f)), different from that of the initial Hamiltonian ($w = 1/2$).  On the contrary, in the short-range sector, the half-way winding number would still be 1 (for instance, the quench protocol in Fig.~\ref{Fig9}(d), see Appendix C), same as that of the original Hamiltonian. Consequently, one can see that the characterization of TPTs via entanglement-spectrum dynamics tightly depends on the difference between the winding number of initial Hamiltonian and the half-way winding number. Moreover, similar to Fig.~\ref{Fig2}(c) and (d), the non-degenerate entanglement-spectrum property can also be explained by the winding number $w=3/2$ (see Appendix D).

It is worthwhile to emphasize that the dynamical behaviors of entanglement spectra close to their crossings are linear in the short-range sector (see Fig~\ref{Fig2}(a)), just like those in the models with nearest-neighbor interactions~\cite{ESnoneq1,ESnoneq2,ESnoneq3}. In stark contrast, as shown in Fig~\ref{Fig2}(c) and (e), the behaviors become non-linear in the long-range sector. We present an explanation of this drastic change in Appendix E.

In addition, we study the above quench protocols with a more detailed method. We fix the initial state as a ground state in the phases with $w=1$, $w=-1/2$ and $w=1/2$, i.e., $(\alpha_{i},\mu_{i}) = (2,3)$, $(0.5,3)$ and $(0.5,-3)$ respectively, and explore the dependence of the $\xi^{(1)}$ at $t=\pi/2$ (denoted as $\xi^{(1)}(t=\pi/2)$) and $\mu_{f}$. In Fig. \ref{Fig3}(a) and (b), with the initial Hamiltonian in the trivial topological phase, $\xi^{(1)}(t=\pi/2)$ shows non-analytical behaviors at the critical points. In Fig. \ref{Fig3}(c), one can see that $\xi^{(1)}(t=\pi/2)$ seems to gradually approach 0.5 as $\mu_f$ gets closer to the critical point $\mu_{c}= 1$. However, at the critical point $\mu_{c}= 1$,
there is a distinctive behavior of $\xi^{(1)}$. In Appendix D, we show that the singularity of the $\xi^{(1)}$ at the critical point is an artifact of band flatting. With the band flatting, the winding number at $t=\pi/2$ is well-defined as the half-way winding number. When $\mu_{f}>\mu_{c}$, the half-way winding number is $w=3/2$. In contrast, if the quench protocol does not cross $\mu_{c}$, the behavior of $\xi^{(1)}$ is similar to the ground-state entanglement spectra in the topological phase of initial Hamiltonian. More specifically, as shown in Fig.~\ref{Fig2}(b), the $\xi^{(1)}$ of ground states also gradually approaches 0.5 with $\mu\rightarrow 1$ when $\alpha=0.5$, $\mu<1$, and the winding number $w=1/2$. Thus, it can be predicted that the behavior of the $\xi^{(1)}$ of ground states for the TPT between the topological phase with $w=1/2$ and $w=3/2$ may be similar to that in Fig. \ref{Fig3}(c). Indeed, we observe a similar singularity of the ground-state $\xi^{(1)}$ near the critical point of a TPT between the $w=1/2$ and $w=3/2$ phase, indicating that the occurrence of the singularity of the ground-state $\xi^{(1)}$ is closely related to the gapless energy spectra (see Appendix C).

%This can be interpreted by considering
%how the half-way winding number changes as $\mu_f$ crosses the critical line. For the quench process in Fig. \ref{Fig3}(a) and (b), there is no massless (or massive) fermion in the phase of initial Hamiltonian ($w=0$ or $-1/2$). However, the half-way winding number is $-2$ (Fig. \ref{Fig2}(b)) or $-3/2$ (Fig. \ref{Fig2}(d)), corresponding to the presence of massless or massive fermion half way through the evolution. Thus, the difference between the initial and final Hamiltonian is similar to that between the Hamiltonian in the phase with $w=0$ and $w=1$, distinguishing the presence or absence of topological edge states. Indeed, as shown in Fig. \ref{Fig1}(b), the ground state of ES $\xi^{(1)}$ behaves nonanalytically at the critical points of the TPTs between the phase with $w=0$ and $w=1$. Employing similar arguement, for the quench process in Fig. \ref{Fig3}(c), the initial Hamiltonian gives rise to a massive topological Driac mode. The half-way winding number is $3/2$, corresponding to one massless edge state half-way through the evolution. Hence, the difference between the initial and final Hamiltonian is similar to that between the Hamiltonian in the phase with $w=1/2$ and $w=1$, which is consistent with the continuous behavior of the ground state ES $\xi^{(1)}$ at the critical point $\alpha_{c}=1$ of the TPT.

\subsection{Decay rate $\alpha$ as the quench parameter}
To further explore the dynamics of entanglement spectra in the long-range system, we now turn to consider the quench protocols with the decay rate $\alpha$ as the quench parameter, i.e., $\alpha=\alpha_{i}\rightarrow \alpha_{f}$, while other parameters are fixed. The $\alpha_{i}$ and $\alpha_{f}$ refer to the decay rate of the initial and final Hamiltonian respectively.

\begin{figure}
	\centering
	\includegraphics[width=1\linewidth]{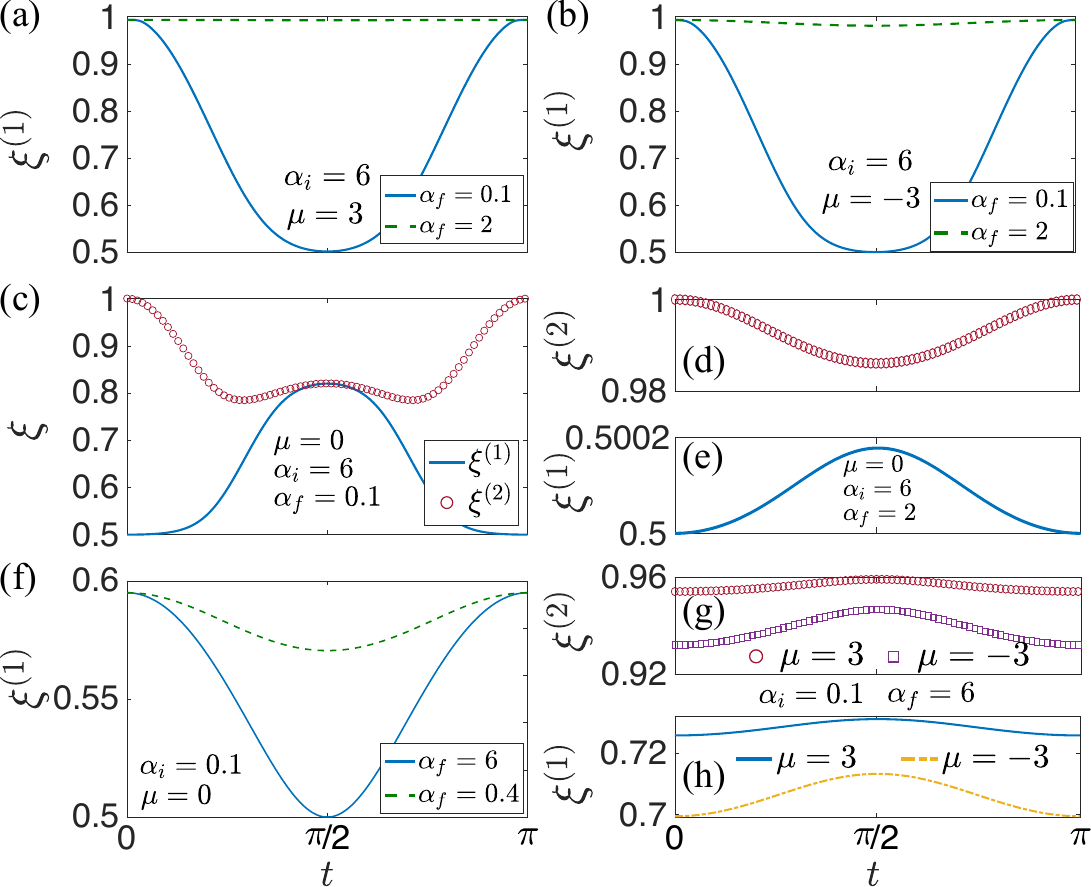}\\
	\caption{(a) Time evolution of the $\xi^{(1)}$ with the quench protocol $\alpha=6\rightarrow 2$ and $0.1$ in the Hamiltonian (\ref{Hamiltonian}) with $\mu=3$ and $\Delta=t=1$. (b) is similar to (a) but in the Hamiltonian (\ref{Hamiltonian}) with $\mu=-3$. (c) Time evolution of the $\xi^{(1)}$ and $\xi^{(2)}$ with the quench protocol $\alpha=6\rightarrow 0.1$ in the Hamiltonian (\ref{Hamiltonian}) with $\mu=0$ and $\Delta=t=1$. (d) and (e) are similar to (c) but with the the quench protocol $\alpha=6\rightarrow 2$. (f) Time evolution of the $\xi^{(1)}$ with the quench protocol $\alpha=0.1\rightarrow 6$ and $0.4$ in the Hamiltonian (\ref{Hamiltonian}) with $\mu=0$ and $\Delta=t=1$. (g) and (h) are similar to (f) but in the Hamiltonian (\ref{Hamiltonian}) with $\mu=\pm 3$.}\label{Fig4}%\end{figure}
\end{figure}

\begin{figure*}
	\centering
	\includegraphics[width=1\linewidth]{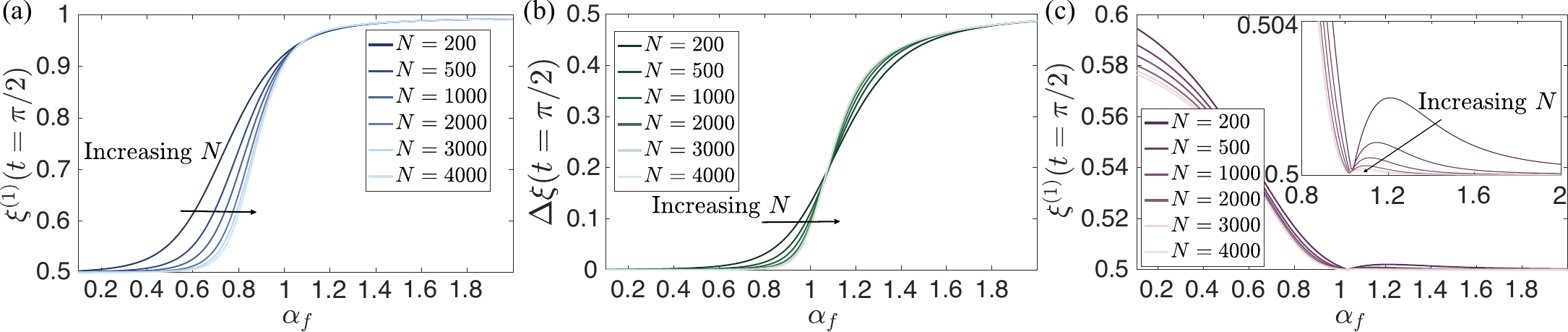}\\
	\caption{(a) The $\xi^{(1)}(t=\pi/2)$, with the quench protocol $\alpha=6\rightarrow \alpha_{f}$ as a function of $\alpha_{f}$ in the Hamiltonian (\ref{Hamiltonian}) with $\mu=3$ and different system sizes $N$. (b) The $\Delta\xi(t=\pi/2)$, with the quench protocol $\alpha=6\rightarrow \alpha_{f}$ as a function of $\alpha_{f}$ in the Hamiltonian (\ref{Hamiltonian}) with $\mu=0$ and different system sizes $N$. Here, the critical points are $\alpha_{c}=1$. (c) is similar to (a) but  with the quench protocol $\alpha=0.1\rightarrow \alpha_{f}$ as a function of $\alpha_{f}$ in the Hamiltonian (\ref{Hamiltonian}) with $\mu=0$. Here, the critical line is $\alpha_{c}=1$.}\label{Fig5}%\end{figure*}
\end{figure*}

Fig. \ref{Fig4} (a)-(e) show the quench dynamics of entanglement spectra with the protocol $\alpha = 6\rightarrow 0.1$ and $2$. For $\mu=3$ ($-3$), the half-way winding number is $+1$ ($-1$) (see Appendix B for the trajectory of $\bm{d}(\pi/2)$) and the entanglement spectra are non-degenerate at $t=\pi/2$. Thus, we only focus on the $\xi^{(1)}$, whose dynamical behaviors are shown in Fig. \ref{Fig4}(a) and (b). The half-way entanglement-spectrum crossings, i.e., $\xi^{(1)}(t=\pi/2) = 0.5$ can still characterize the TPTs with the critical line $\alpha_{c}=1$. Next, we study the quench protocol $\alpha = 6\rightarrow 0.1$ with $\mu=0$. Different from the previous protocols where the ground-state $\xi^{(1)}$ of $H^{i}$ is larger than that of $H^{f}$, such as the quench protocols in Fig. \ref{Fig4}(a) and (b), the ground-state $\xi^{(1)}$ of $H^{i}$ in the current protocol is smaller than that of $H^{f}$. In fact, the $\xi^{(1)}$ of $H^{i}$ is equal to $0.5$, as depicted in Fig.~\ref{Fig1}(b). Remarkably, as shown in Fig. \ref{Fig4}(c), (d) and (e), instead of the entanglement-spectrum crossings $\xi=1/2$, the TPTs can be alternatively characterized by a novel behavior of the entanglement spectra, i.e., $\xi^{(1)}(t=\pi/2)=\xi^{(2)}(t=\pi/2)$ when quenching across the critical line $\alpha_{c}=1$ ($\alpha_{f}=0.1$), while there is a large discrepancy between $\xi^{(1)}(t=\pi/2)$ and $\xi^{(2)}(t=\pi/2)$ when staying the same phase ($\alpha_{f}=2$).

In addition, we also study the inverse of above protocols. Fig.~\ref{Fig4} (f)-(h) show the quench dynamics of entanglement spectra with the protocol $\alpha = 0.1\rightarrow 6$ and $0.4$. For $\mu=0$, the ground-state $\xi^{(1)}$ of $H^{i}$ is larger than that of $H^{f}$, and the dynamical entanglement-spectrum properties are similar to Fig. \ref{Fig4}(a) and (b). For $\mu=\pm 3$, even if the quenches cross the critical line $\alpha_{c}=1$, $\xi^{(1)}(t=\pi/2)=\xi^{(2)}(t=\pi/2)$ is not satisfied, and the ES dynamics fails to characterize the TPTs. Actually, in comparison with the quench protocol in Fig.~\ref{Fig4} (c) where the half-way winding number ($w=0$) is different from that of initial Hamiltonian ($w=1$), for the quench protocol in Fig.~\ref{Fig4} (g) and (h), the dynamics of entanglement spectra is predicted to be trivial since the winding number of initial Hamiltonian is equal to the half-way winding number. Complementary with previous results in Fig.~\ref{Fig2} and Fig.~\ref{Fig9}(d), one can see that the difference between half-way winding number and the winding number of initial Hamiltonian is tightly related to the capability of entanglement-spectrum dynamics in identifying TPTs.

We then fix the initial Hamiltonian with the parameters $(\alpha_{i},\mu_{i}) = (6,3)$, $(6,0)$ and $(0.1,0)$, and explore the entanglement spectra at $t=\pi/2$ as a function of $\alpha_{f}$. In Fig.~\ref{Fig5}(a), the dependence of the $\xi^{(1)}(t=\pi/2)$ and $\alpha_{f}$ with the quench protocols $\alpha=6\rightarrow \alpha_{f}$ and $\mu=3$ is presented. One can see that the results of system size $N=200$ suffers from finite-size effect. Thus, we further calculate the results of larger system size $N=500-4000$. When increasing $N$, the change of $\xi^{(1)}(t=\pi/2)$ becomes more dramatic at the critical point $\alpha_{c}=1$. To study another quench protocols $\alpha=6\rightarrow \alpha_{f}$ with $\mu=0$, we focus on the difference between the first and second lowest entanglement-spectrum value at $t=\pi/2$, i.e., $\Delta\xi = \xi^{(2)} - \xi^{(1)}$. The difference of the entanglement spectra $\Delta\xi(t=\pi/2)$ as a function of $\alpha_{f}$ is shown in Fig.~\ref{Fig5}(b). Similar to the results in Fig.~\ref{Fig5}(a), with the increase of $N$, the entanglement-spectrum critical behavior becomes more obvious. In Fig.~\ref{Fig5}(c), we present the results of the quench protocols $\alpha=0.1\rightarrow \alpha_{f}$ with $\mu=0$, as the quenches with opposite direction of these in Fig.~\ref{Fig5}(b). On this condition, the finite-size effect is smaller, and it can be directly inferred that the $\xi^{(1)}(t=\pi/2)$ has finite value for $\alpha_{f}<\alpha_{c}$ while vanishes for $\alpha_{f}>\alpha_{c}$ ($\alpha_{c}=1$) when $N\rightarrow \infty$.

\begin{figure}
	\centering
	\includegraphics[width=1\linewidth]{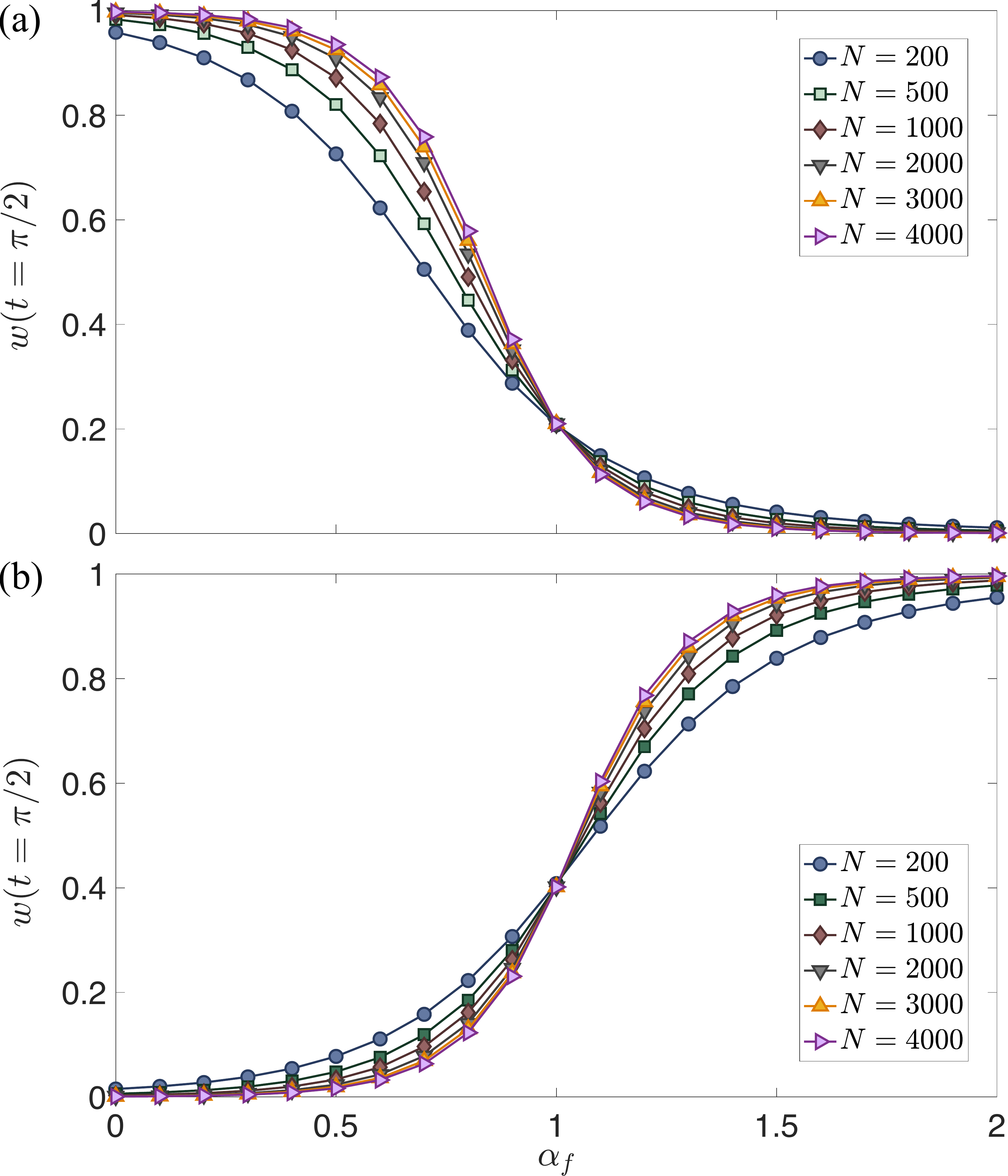}\\
	\caption{(a) The half-way winding number $w(t=\pi/2)$ as a function of $\alpha_{f}$ with the quench protocol in Fig.~\ref{Fig5}(a).  (b) is similar to (a) but with the quench protocol in Fig.~\ref{Fig5}(b). }\label{Fig_add1}%\end{figure}
\end{figure}

It is noted that the finite-size effects in long-range systems may be peculiar, and the crossing point of all the curves may not necessarily correspond to the critical point (an example is given in Ref.~\cite{Dicke}). In Fig.~\ref{Fig5}(a) and (b), the results of $\xi^{(1)}(t=\pi/2)$ show a strong finite-size effect. To better understand the finite-size effect in long-range sector, in Fig.~\ref{Fig_add1}), we plot the half-way winding number $w(t=\pi/2)$ as a function of $\alpha_{f}$ with the same quench protocol in Fig.~\ref{Fig5}(a) and (b), showing that the finite-size behaviors of $w(t=\pi/2)$ are similar to these of $\xi^{(1)}(t=\pi/2)$.

\section{Summary and outlook}
Recent works \cite{ESnoneq1,ESnoneq2,ESnoneq3} show that the TPTs in the systems with nearest-neighbor interactions can be characterized by the quench dynamics of entanglement spectra. By studying the out-of-equilibrium properties of entanglement spectra in the long-range Kitaev chain with power-law decay pairing terms, we have found that: (i) in the short-range sector (decay rate $\alpha>1$), the entanglement-spectrum behaviors are similar to these in the conventional Kitaev chain~\cite{ESnoneq3}, i.e., when quenching across the critical line and the initial Hamiltonian is topologically trivial, the entanglement-spectrum crossings $\xi=1/2$ are observed. (ii) In the long-range sector (decay rate $\alpha<1$), for both topologically trivial or non-trivial initial Hamiltonian, the entanglement-spectrum crossings can still characterize the TPTs via entanglement-spectrum crossing. However, the entanglement-spectrum crossings vanish for the topologically trivial final Hamiltonian in the short-range sector even if the quench protocols cross the critical lines. (iii) For the quench protocols with decay rate $\alpha$ as the quench parameter, the entanglement-spectrum dynamics can diagnose TPTs when the winding number of initial Hamiltonian is different from the half-way winding number. Besides the entanglement-spectrum crossings, one can also characterize the TPTs by studying the difference between the first and second lowest entanglement-spectrum value in the special case. In a word, the characterization of TPTs via the quench dynamics of entanglement spectra could be well generalized to long-range systems. Moreover, we emphasize that although the following results are based on the band-flattened Hamiltonian, as shown in Appendix D, the TPTs can still be characterized by the entanglement-spectrum crossings for the post-quench Hamiltonian without flat bands.

This work may inspire further investigations on the quench dynamics of entanglement spectra in several long-range systems, for instance, the characterization of the topological phases with higher winding number in the longer-range Kitaev chains~\cite{high_winding1,high_winding2}, the TPTs in the two-dimensional topological superconductors with long-range interactions~\cite{two_dim}, the conventional quantum phase transitions~\cite{CQPT} or dynamical phase transitions~\cite{DQPT,DQPT_exp,DQPT_add,DQPT_add2} in long-range systems.

\begin{acknowledgments}
We acknowledge discussions with Zongping Gong, Jinlong Yu, Shuangyuan Lu, and N. Sedlmayr. This work was supported by the National Natural Science Foundation of China (11934018), National Basic Research Program of China (2016YFA0302104, 2016YFA0300600), and Strategic Priority Research Program of Chinese Academy of Sciences (XDB28000000).
\end{acknowledgments}

\appendix
\section{The antiperiodic boundary condition}
For the long-range Kitaev chain with power-law decay pairing terms, we adopt the antiperiodic boundary condition $c_{i+N}=-c_{i}$. Here, we discuss the reason behind the choice of the antiperiodic boundary condition in detail.

In the Hamiltonian (\ref{Hamiltonian}), the term describing the power-law decay pairing interactions can be regrouped, i.e., $\sum_{i=1}^{N}\sum_{l=1}^{N-i}\frac{1}{d_{l}^{\alpha}}(c_{i}c_{i+l} + \text{H.c.})=\sum_{i=1}^{N}\sum_{i=1}^{N-l}\frac{1}{d_{l}^{\alpha}}(c_{i}c_{i+l} + \text{H.c.})$. Then, we can pick the $l$ nearest terms $\sum_{i=1}^{N-l}\frac{1}{d_{l}^{\alpha}}(c_{i}c_{i+l} + \text{H.c.})$ and $N-l$ nearest $\sum_{i=1}^{l}\frac{1}{d_{N-l}^{\alpha}}(c_{i}c_{i+N-l} + \text{H.c.})$. We note that the coefficients of the $l$ and $N-l$ nearest terms are actually the same since $d_{l}=d_{N-l}$. As a consequence, taking $l=1$ for example, we can see $c_{1}c_{2}+\ldots+c_{N-1}c_{N} + (c_{1}c_{N}) = c_{1}c_{2}+\ldots+c_{N-1}c_{N} - c_{1}c_{N}$. Next, imposing the antiperiodic boundary condition $c_{i+N}=-c_{i}$, we have $c_{1}c_{2}+\ldots+c_{N-1}c_{N} + c_{N}c_{N+1}$, which allows us to switch the Hamiltonian to the Fourier basis. If the periodic boundary condition is imposed, the coefficients of the $l$ and $N-l$ nearest terms should be the opposite of each other.

\section{The details of calculating the quench dynamics of entanglement spectra}
After obtaining the winding vector at arbitary time $t$ during the quench dynamics according to Eq. (\ref{dynamic2}), we can diagonalize the Hamiltonian at $t$, i.e., $H(t)=\sum_{k}\epsilon_{k}\Psi_{k}^{\dagger}(\bm{d}_{k}(t)\cdot \bm{\sigma})\Psi_{k}$ , via the Bogoliubov transformation
\begin{eqnarray}
\Psi_{k}^{\dagger}(\bm{d}_{k}(t)\cdot \bm{\sigma})\Psi_{k} = \Psi_{k}^{\dagger}V_{k}(t) \Lambda_{k}(t) V_{k}^{\dagger}(t)\Psi_{k},
\label{bogoliubov}
\end{eqnarray}
giving the Bogoliubov fermion operators at $t$, i.e., $(a_{-k}(t), a_{k}^{\dagger}(t)) = \Psi_{k}^{\dagger}V_{k}(t)$, which satisfies $a_{k}(t)| \text{Vac} \rangle $ =0. The two-dimension unitary matrix $V_{k}$, whose elements are denoted as $v_{ij}(k)$ $(i,j\in\{1,2\})$, is the key to calculate the correlation matrix
\begin{eqnarray}
C_{mn} &=& \langle \text{Vac} | c_{m}^{\dagger}c_{n} | \text{Vac} \rangle \nonumber \\
&=& \frac{1}{N} \sum_{k} e^{-ik(m-n)} v_{22}(k) v^{*}_{22}(k) \nonumber \\
&+&e^{ik(m-n)} v_{11}(k) v^{*}_{11}(k),
\end{eqnarray}
and anomalous correlation matrix,
\begin{eqnarray}
F_{mn} &=& \langle \text{Vac} | c_{m}^{\dagger}c_{n}^{\dagger} | \text{Vac} \rangle \nonumber \\
&=& \frac{1}{N} \sum_{k} e^{-ik(m-n)} v_{22}(k) v^{*}_{12}(k) \nonumber \\
&+&e^{ik(m-n)} v_{21}(k) v^{*}_{11}(k),
\end{eqnarray}
which are useful for obtaining the entanglement spectra.

The value of $\Omega_{q}$ in Eq. (\ref{ES_def}) can be obtained by the diagonalization of a matrix composed of the correlation matrix and anomalous correlation matrix, i.e.,
\begin{eqnarray}
\begin{pmatrix} I-C_{\mathcal{A}}^{*} & -F_{\mathcal{A}}^{*} \\ F_{\mathcal{A}} & C_{\mathcal{A}} \end{pmatrix}
= P^{\dagger}
\begin{pmatrix} \Xi^{-} & 0 \\ 0 & \Xi^{+} \end{pmatrix}  P,
\end{eqnarray}
with $\Xi^{\pm}$ as diagonal matrix whose elements are $1/(1+e^{\pm \Omega_{q}})$.
\begin{figure}
	\centering
	\includegraphics[width=1\linewidth]{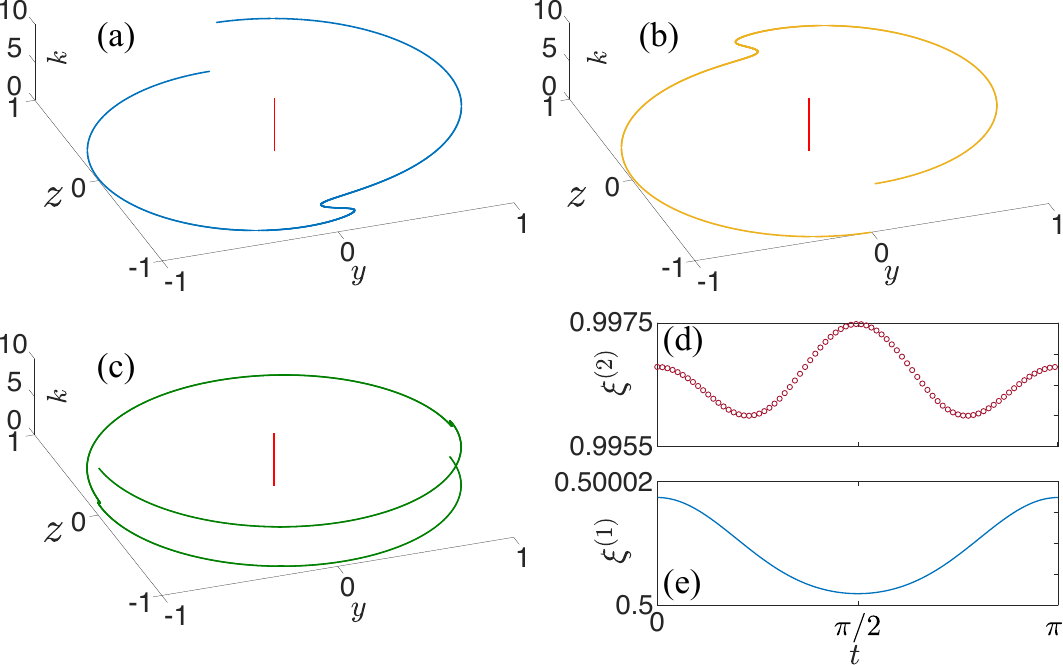}\\
	\caption{(a) The trajectory of winding vector at time $t=\pi/2$ for the quench protocol in Fig.~\ref{Fig4}(a) with $\alpha_{f}=0.1$. (b) is similar to (a) but with the protocol in Fig.~\ref{Fig4}(b) with $\alpha_{f}=0.1$. (c) is similar to (a) but with the protocol in Fig.~\ref{Fig4}(f) with $\alpha_{f}=6$. (d) and (e) show the time evolution of the $\xi^{(2)}$ and $\xi^{(1)}$ with the quench protocol $\mu=0.5 \rightarrow 3$ and other parameters are similar to the quench protocol in Fig.~\ref{Fig2}(a). }\label{Fig9}%\end{figure*}
\end{figure}
\section{Additional results}
In this appendix, we present some addition results, including the trajectory of winding vector at time $t=\pi/2$ and time evolution of the entanglement spectra for some quench protocols.

\section{Extended Kitaev chains with long-range pairing and the topological phase with $w=3/2$}
\begin{figure}
	\centering
	\includegraphics[width=1\linewidth]{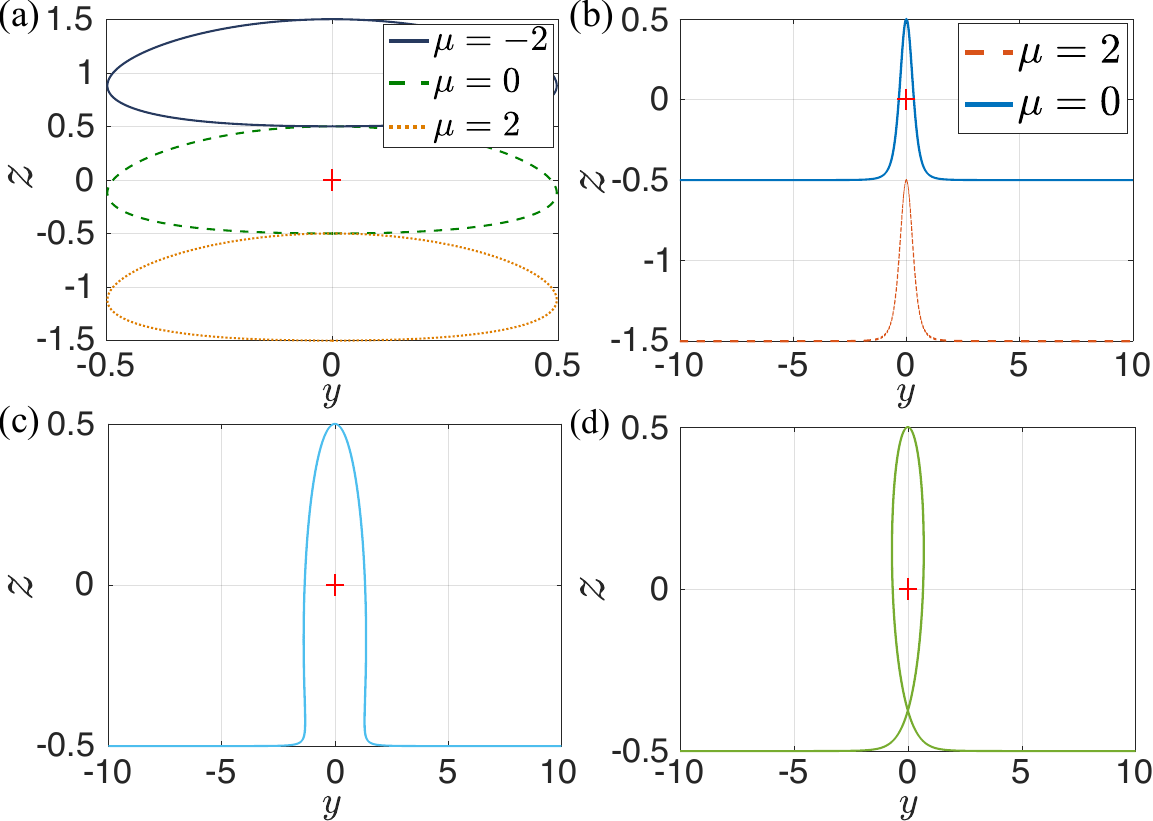}\\
	\caption{The trajectory of winding vector in the Hamiltonian (\ref{Hamiltonian_toy}) with (a) $\alpha=2$, $\Delta=t=1$, $\delta=0$, and $\mu=-2$, $0$, $2$. The winding number $w=1$ for $\mu=0$ and $w=0$ for $\mu=\pm 2$. (b) $\alpha=0.5$, $\Delta=t=1$, $\delta=0$, and $\mu=0$, $2$. The winding number $w=1/2$ for $\mu=0$ and $w=-1/2$ for $\mu=2$. (c) $\mu=0$, $\alpha=0.5$, $t=\Delta=1$, and $\delta=2$. The winding number $w=1/2$. (d) $\mu=0$, $\alpha=0.5$, $t=\Delta=1$, and $\delta=-2$. The winding number $w=3/2$.}\label{Fig7}%\end{figure*}
\end{figure}

\begin{figure}
	\centering
	\includegraphics[width=0.8\linewidth]{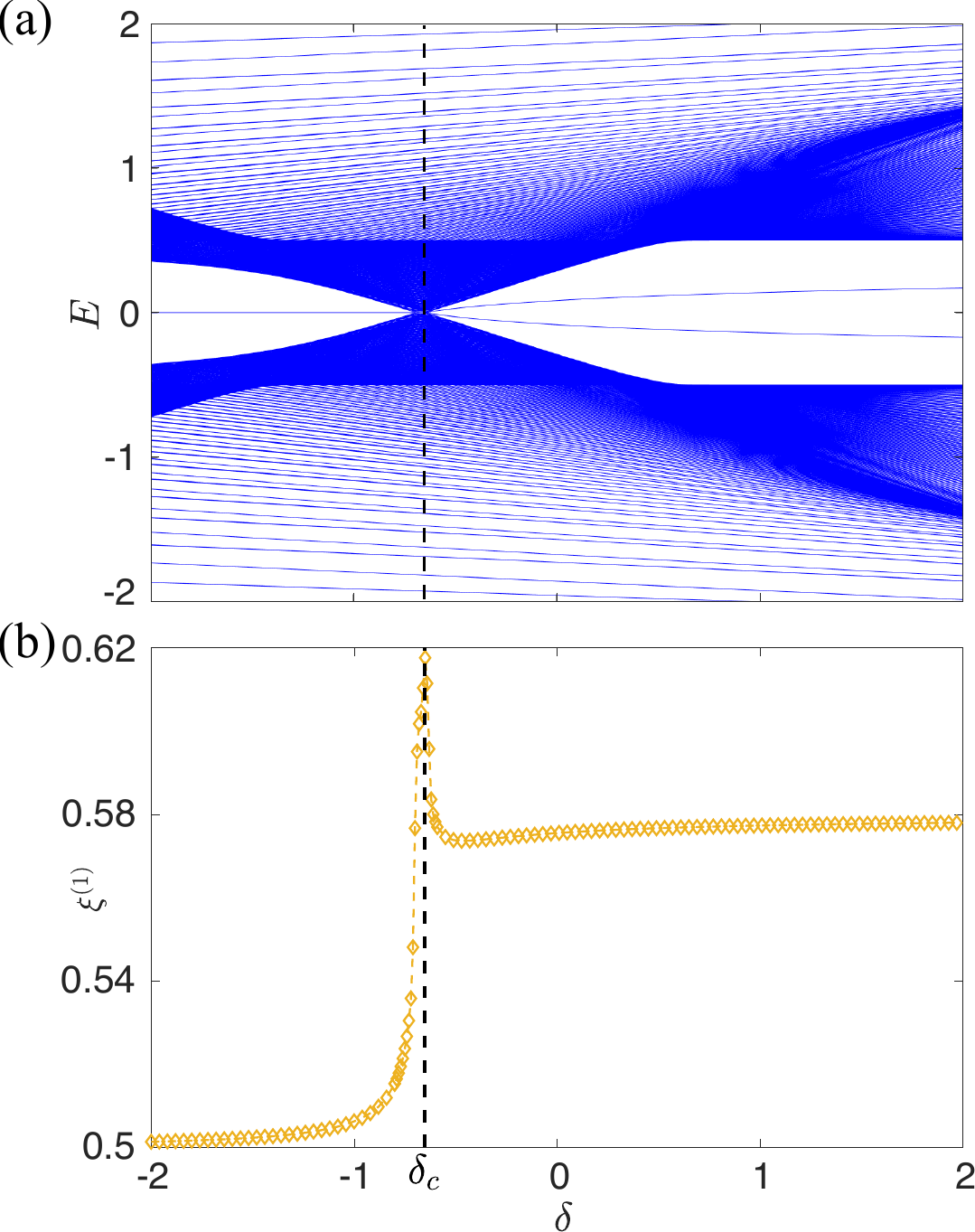}\\
	\caption{(a) The energy spectrum for the Hamiltonian (\ref{Hamiltonian_toy}) as a function of $\delta$. (b) The $\xi^{(1)}$ as a function of $\delta$ for the ground state of the Hamiltonian (\ref{Hamiltonian_toy}). Here, the other parameters are $\mu=0$, $\alpha=0.5$, and $t=\Delta=1$. }\label{Fig8}%\end{figure*}
\end{figure}

Here, we consider a new Hamiltonian with the open boundary condition
\begin{eqnarray}
H = &-&\frac{t}{2}\sum_{i=1}^{N}(c_{i}^{\dagger}c_{i+1}+\text{H.c.}) - \mu\sum_{i=1}^{N}(c_{i}^{\dagger}c_{i}-\frac{1}{2}) \nonumber \\
&+& \frac{\Delta}{2}\sum_{i=1}^{N-l}\sum_{l=1}^{N/2-1}\frac{1}{d_{l}^{\alpha}}(c_{i}c_{i+l} + \text{H.c.}) \nonumber \\
&+& \frac{\delta}{2}\sum_{i=1}^{N-1}(c_{i}c_{i+1} + \text{H.c.}).
\label{Hamiltonian_toy}
\end{eqnarray}
It is noted that the summation notation of the long-range pairing term $\frac{1}{d_{l}^{\alpha}}(c_{i}c_{i+l} + \text{H.c.})$ is sightly different from the one in Eq. (\ref{Hamiltonian}) because of the cutoff of the power-law decay $N/2-1$ for convenience. The winding vector of the Hamiltonian (\ref{Hamiltonian_toy}) reads
\begin{eqnarray}
\bm{d}_{k} &=& (d_{k}^{x},d_{k}^{y},d_{k}^{z}) \nonumber \\
&=& (0,-\frac{(\Delta f_{\alpha}(k)+\delta\sin k)}{2},-\frac{(\mu+t\cos k)}{2}).
\label{Anderson_toy}
\end{eqnarray}

Then, we plot the the trajectory of the winding vector, obtaining the related winding number. In Fig. \ref{Fig7}(a) and (b), we show that with $\delta=0$, the winding numbers of the topological phases in Hamiltonian (\ref{Hamiltonian_toy}) are the same as these in Hamiltonian (\ref{Hamiltonian}). Thus, the cutoff of the power-law decay $N/2-1$ does not influence the phase diagram of the Hamiltonian. More importantly, with $\mu=0$, $\alpha=0.5$, and $t=\Delta=1$, as shown in Fig. \ref{Fig7}(c) and (d), the winding number is equal to $1/2$ and $3/2$ for $\delta=2$ and $-2$ respectively. Actually, by tuning $\delta$ from $2$ to $-2$, there is a TPT between the topological phases with $w=1/2$ and $w=3/2$.

In addition, we also present the energy spectrum as a function of $\delta=2$ with $\mu=0$, $\alpha=0.5$, and $t=\Delta=1$ in Fig.~\ref{Fig8}(a). One can see that there are massive Dirac edge states in the topological phase with $w=1/2$ (for instance, $\delta=2$), and one pair of Majorana zero modes is present in the topological phase with $w=3/2$ (for instance, $\delta=-2$). We also calculate the $\xi^{(1)}$ for the ground state, which is plotted in Fig.~\ref{Fig8}(b). For $\delta<\delta_{c}$, $\xi^{(1)}\simeq 0.5$, corresponding to the one pair of Majorana zero modes, while $\xi^{(1)}$ has a larger value when $\delta>\delta_{c}$ and massive Dirac edge states are observed. There is a singularity of $\xi^{(1)}$ close to the critical point $\delta_{c}$, which is connected with the gapless energy spectrum. In contrast, as shown in Fig.~\ref{Fig1}(b), the singularity of $\xi^{(1)}$ is absent at the critical line $\alpha_{c}=1$ since the energy spectrum is gapped (see Fig.~\ref{Fig1}(c) and (d)).

It is noted that with the same parameters of the topological phase $w=3/2$ in the Hamiltonian (\ref{Hamiltonian_toy}), after performing a gauge transformation $c_{j}\rightarrow c_{j}e^{i\frac{\pi}{2}}$, the winding number becomes $w=-3/2$. Since the gauge transformation does not influence the topological properties of the system, it is predicted that there is also one pair of Majorana zero modes in the topological phase with $w=-3/2$, which corresponds to the entanglement-spectrum non-degeneracy in Fig.~\ref{Fig2}(e).

\section{An explanation of the non-linear behaviors of $\xi^{(1)}$ near $t=\pi/2$}
In this Appendix, we present a perturbation study to explain the linear and non-linear behaviors of $\xi^{(1)}$ around $t=\pi/2$ in the short-range and long-range sector respectively. According to Eq.~(\ref{ES_energy}), $\xi_{q}\propto \lambda_{q}$, and thus we can perform a perturbation study of the eigenenergies to understand the entanglement-spectrum properties.

From Eq.(\ref{dynamic2}), only keeping the linear term, we have
\begin{eqnarray}
\Delta\bm{d}_{k} &=& \bm{d}_{k}(\pi/2+\Delta t) - \bm{d}_{k}(\pi/2) \nonumber \\
&=& -2\Delta t [\bm{d}_{k}^{i}\times\bm{n}_{k}^{f}].
\label{delta_dk}
\end{eqnarray}
Thus, the perturbation problem now reduces to perturbing the half-way Hamiltonian $H_{0}$ with $-2\Delta t V$ with $V$ as the Hamiltonian corresponding to $\bm{d}_{k}^{i}\times\bm{n}_{k}^{f}$. It is seen that the perturbation term preserves the particle-hole symmetry, but breaks the time reversal symmetry, i.e., $\mathcal{P}V\mathcal{P}^{-1}=-V$ and $\mathcal{T}V\mathcal{T}^{-1}=-V$.

We now focus on the lowest two eigenstates of the half-way Hamiltonian, and denote the eigenmodes as $c_{1}$ and $c_{2}$. The couplings between $V$ and the two eigenmodes can be described by a four-level effective model $V = \psi^{\dagger} V_{\text{eff}} \psi$, where $\psi^{\dagger} = (c_1^\dagger,  c_2^\dagger,  c_1,  c_2)$, and
\begin{align}
	V_{\text{eff}} = \begin{pmatrix}
	A & B \\
	-B^* & -A^* \\
	\end{pmatrix}
	\end{align}
with $A$ and $B$ as 2-dimensional matrix. The above structure is a requirement of particle-hole symmetry. Then, considering $\mathcal{T}V\mathcal{T}^{-1}=-V$, We recognize that $A=-A^{*}$ and $B=-B^{*}$. Moreover, because of the anti-communication relation, the matrix $A$ is eliminated and $B$ should be anti-symmetric. Finally, the form of $V_{\text{eff}}$ can be fixed as
\begin{align*}
	V_{\text{eff}}\propto\begin{pmatrix}
	0& 0& 0& i \\
	0& 0& -i& 0 \\
	0& i& 0& 0 \\
	-i& 0& 0& 0
	\end{pmatrix},
	\end{align*}
i.e., $V_{\text{eff}}\propto i(c_{2}c_{1}-c_{1}^{\dagger}c_{2}^{\dagger})$.

In the short-range sector, the zero energy modes are degenerate, and can be decomposed into Majorana operators residing on the left and right end of the chain. Specifically, we have $c_{i}=a_{i}+ib_{i}$ ($i=1,2$), and $V_{\text{eff}}\propto -2i(a_{1}a_{2}+b_{2}b_{1})$. It directly couples Majorana modes on the same side, giving a linear contribution to the energy.

In the long-range sector, $c_{1}$ is still massless, while $c_{2}$ becomes massive. Here,
\begin{align*}
	H_{0}-2\Delta tV_{\text{eff}}\propto\begin{pmatrix}
	0& 0& 0& i\Delta t' \\
	0& \epsilon& -i\Delta t'& 0 \\
	0& i\Delta t'& 0& 0 \\
	-i\Delta t'& 0& 0& -\epsilon
	\end{pmatrix}
	\end{align*}
with $\epsilon\neq 0$ since $c_{2}$ denotes a massive mode, and $\Delta t'$ as a normalized perturbation parameter proportional to $\Delta t$. The eigenvalues of the matrix are
\begin{align}
	\lambda^{(1)}_{\pm}=\sqrt{(\frac{\epsilon}{2})^{2} + (\Delta t')^{2}}\pm\frac{\epsilon}{2} \\ \nonumber
 \lambda^{(2)}_{\pm}=-\sqrt{(\frac{\epsilon}{2})^{2} + (\Delta t')^{2}}\pm\frac{\epsilon}{2}. \nonumber
\end{align}
We note that $V$ also contains other terms that couple $c_{1}$ and $c_{2}$ to other bulk modes. Nevertheless, these terms do not give any contributions in the leading order, and thus do not influence the above analysis. Since we have shown that in the long-range sector, the eigenvalues can be regarded as non-linear functions of $\Delta t$, the above discussion explains the difference between the behaviors of $\xi^{(1)}$ near $t=\pi/2$ in the short-range and long-range sector.

\begin{figure*}[]
	\centering
	\includegraphics[width=1\linewidth]{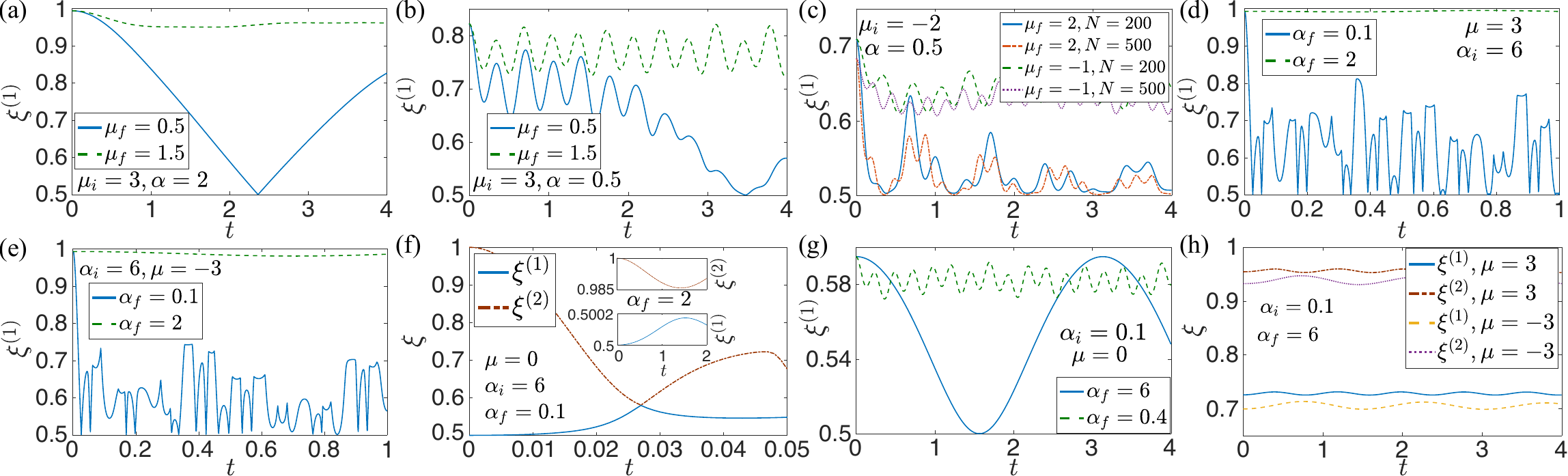}\\
	\caption{The quench dynamics of entanglement spectra for the post-quench Hamiltonian without flat bands. The quanch protocol in (a)-(e) is the same as that in Fig.~\ref{Fig2}(a), Fig.~\ref{Fig2}(c), Fig.~\ref{Fig2}(e), Fig.~\ref{Fig4}(a) and Fig.~\ref{Fig4}(b), respectively. The quench protocol in (f) is the same as that in Fig.~\ref{Fig4}(c)-(e). The quench protocol in (g) is the same as that in Fig.~\ref{Fig4}(f). The quench protocol in (h) is the same as that in Fig.~\ref{Fig4}(g) and (h). }\label{Fig6}%\end{figure*}
\end{figure*}

\begin{figure}
	\centering
	\includegraphics[width=1\linewidth]{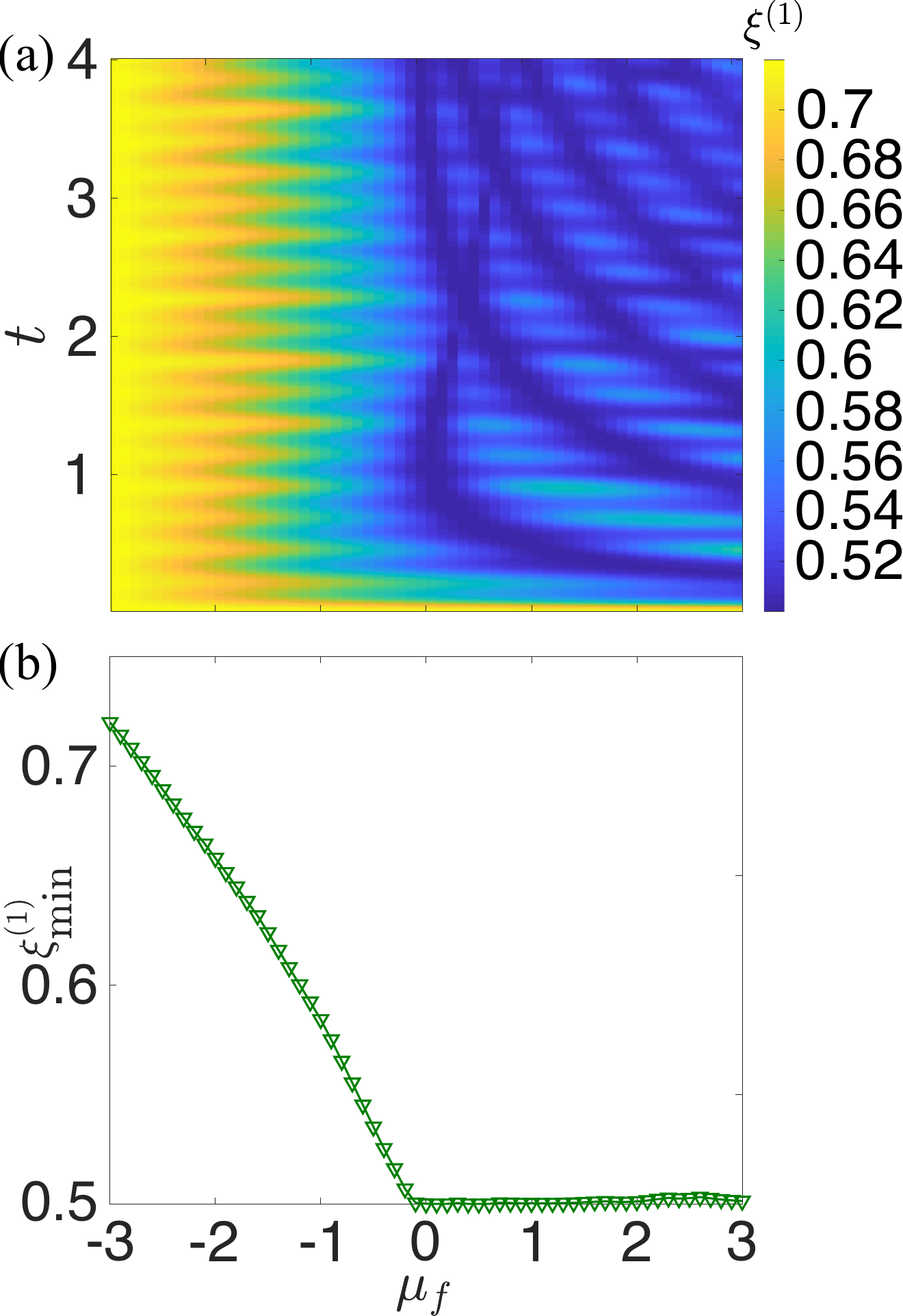}\\
	\caption{(a) The $\xi^{(1)}$ plotted in the $t$-$\mu_{f}$ plane. (b) The $\xi^{(1)}_{\text{min}}$ as a function of $\mu_{f}$.  }\label{Fig10}%\end{figure*}
\end{figure}

\section{Results of the post-quench Hamiltonian without flat bands}
In this Appendix, we present the quench dynamics of entanglement spectra with the same quench protocols in Fig.~\ref{Fig2}(a), (c), (e) and Fig.~\ref{Fig4}. The results are shown in Fig~\ref{Fig6}, indicating that the conclusions made in the flat-band case are stable. Indeed, it has been shown that the occurrence of entanglement-spectrum crossings is stable for the conventional Kitaev chain with nearest interactions in class D~\cite{ESnoneq3}. Here, we demonstrate that the stability of the crossings can be generalized to the long-range Kitaev chain.

It is noted that a distinctive fast oscillating entanglement-spectrum behavior can be observed when the initial or finial Hamiltonian is in the long-range sector. For instance, the oscillation of entanglement spectra is more dramatic in Fig.~\ref{Fig6}(b) than that in Fig.~\ref{Fig6}(a). The oscillating behavior may be related to the divergence of quasiparticle energy since for the Hamiltonian (\ref{Hamiltonian}), the divergence of $\epsilon_{k}$ can occur when $k=0$ or $2\pi$ in the long-range sector $\alpha<1$~\cite{Ek}.

Furthermore, we study the quench dynamics of $\xi^{(1)}$ without flat bands in detail. The initial state is chosen as a ground state of Hamiltonian (\ref{Hamiltonian}) with $(\alpha_{i},\mu_{i})=(0.5,-3)$, and $\mu_{f}$ varies from -3 to 3. In Fig.~\ref{Fig10}(a), it is shown that although $\mu_{f}<\mu_{c}$ ($\mu_{c}=1$), the $\xi^{(1)}$ approaches 0.5 during its time evolution. Moreover, we can employ the minimum entanglement-spectrum value  $\xi^{(1)}_{\text{min}}=\min_{t\in[0,t_{f}]} \xi^{(1)}(t)$ with ($t_{f}=4$) to clarify whether the quench dynamics of entanglement spectra can efficiently detect the TPT without flat bands. The $\xi^{(1)}_{\text{min}}$ as a function of $\mu_{f}$ is plotted in Fig.~\ref{Fig10}(b). It is noted that the quench protocol in Fig.~\ref{Fig10} is the same as that in Fig.~\ref{Fig3}(c). However, the entanglement-spectrum singular behavior near the critical point vanishes without flat bands.

Here, we emphasize that although the qualitative behaviors of the entanglement spectra are stable against energy dispersion (see Fig.~\ref{Fig6}), in Fig.~\ref{Fig10}(b), the $\xi^{(1)}_{\text{min}}$ is trivial around the critical point $\mu_{c}=1$. As a consequence, the characterization of TPTs via the entanglement-spectrum dynamics in the long-range system without flat bands remains an open and complex problem, and deserves a more special study.
%\section{Topological phase with winding number $w=3/2$}

%We add a term of nearest pairing interaction to the Hamiltonian $H$ in  Eq. (\ref{Hamiltonian}), constructing
%\begin{eqnarray}
%H_{2} = H + \frac{\Delta_{2}}{2}\sum_{i=1}^{N-1}(c_{i}c_{i+1} + \text{H.c.}).
%\label{Hamiltonian2}
%\end{eqnarray}
%The winding vector of $H_{2}$ can be directly obtained, i.e.,
%$\bm{d}_{k} = (0,d_{k}^{y},d_{k}^{z})$ with

%\begin{eqnarray}
%d_{k}^{y} &=& -0.5(\Delta f_{\alpha}(k)-\Delta_{2}\sin k), \nonumber \\
%d_{k}^{z} &=& -(\mu+t\cos k)
%\label{Anderson2}
%\end{eqnarray}
%In Fig. , it is revealed that there is a TPT between the topological phase with $w=1/2$ and $w=3/2$ by varying the parameter $\Delta_{2}$. From the energy spectrum in Fig. , one can see that the massive edge modes is turned back into the massless one when changing $\Delta_{2}$ from $2$ to $-2$.

\bibliography{reference}

\begin{thebibliography}{66}%
\makeatletter
\providecommand \@ifxundefined [1]{%
 \@ifx{#1\undefined}
}%
\providecommand \@ifnum [1]{%
 \ifnum #1\expandafter \@firstoftwo
 \else \expandafter \@secondoftwo
 \fi
}%
\providecommand \@ifx [1]{%
 \ifx #1\expandafter \@firstoftwo
 \else \expandafter \@secondoftwo
 \fi
}%
\providecommand \natexlab [1]{#1}%
\providecommand \enquote  [1]{``#1''}%
\providecommand \bibnamefont  [1]{#1}%
\providecommand \bibfnamefont [1]{#1}%
\providecommand \citenamefont [1]{#1}%
\providecommand \href@noop [0]{\@secondoftwo}%
\providecommand \href [0]{\begingroup \@sanitize@url \@href}%
\providecommand \@href[1]{\@@startlink{#1}\@@href}%
\providecommand \@@href[1]{\endgroup#1\@@endlink}%
\providecommand \@sanitize@url [0]{\catcode `\\12\catcode `\$12\catcode
  `\&12\catcode `\#12\catcode `\^12\catcode `\_12\catcode `\%12\relax}%
\providecommand \@@startlink[1]{}%
\providecommand \@@endlink[0]{}%
\providecommand \url  [0]{\begingroup\@sanitize@url \@url }%
\providecommand \@url [1]{\endgroup\@href {#1}{\urlprefix }}%
\providecommand \urlprefix  [0]{URL }%
\providecommand \Eprint [0]{\href }%
\providecommand \doibase [0]{http://dx.doi.org/}%
\providecommand \selectlanguage [0]{\@gobble}%
\providecommand \bibinfo  [0]{\@secondoftwo}%
\providecommand \bibfield  [0]{\@secondoftwo}%
\providecommand \translation [1]{[#1]}%
\providecommand \BibitemOpen [0]{}%
\providecommand \bibitemStop [0]{}%
\providecommand \bibitemNoStop [0]{.\EOS\space}%
\providecommand \EOS [0]{\spacefactor3000\relax}%
\providecommand \BibitemShut  [1]{\csname bibitem#1\endcsname}%
\let\auto@bib@innerbib\@empty
%</preamble>
\bibitem [{\citenamefont {Qi}\ and\ \citenamefont {Zhang}(2011)}]{rev1}%
  \BibitemOpen
  \bibfield  {author} {\bibinfo {author} {\bibfnamefont {X.-L.}\ \bibnamefont
  {Qi}}\ and\ \bibinfo {author} {\bibfnamefont {S.-C.}\ \bibnamefont {Zhang}},\
  }\href {\doibase 10.1103/RevModPhys.83.1057} {\bibfield  {journal} {\bibinfo
  {journal} {Rev. Mod. Phys.}\ }\textbf {\bibinfo {volume} {83}},\ \bibinfo
  {pages} {1057} (\bibinfo {year} {2011})}\BibitemShut {NoStop}%
\bibitem [{\citenamefont {Sato}\ and\ \citenamefont {Ando}(2017)}]{rev2}%
  \BibitemOpen
  \bibfield  {author} {\bibinfo {author} {\bibfnamefont {M.}~\bibnamefont
  {Sato}}\ and\ \bibinfo {author} {\bibfnamefont {Y.}~\bibnamefont {Ando}},\
  }\href {\doibase 10.1088/1361-6633/aa6ac7} {\bibfield  {journal} {\bibinfo
  {journal} {Reports on Progress in Physics}\ }\textbf {\bibinfo {volume}
  {80}},\ \bibinfo {pages} {076501} (\bibinfo {year} {2017})}\BibitemShut
  {NoStop}%
\bibitem [{\citenamefont {Nadj-Perge}\ \emph {et~al.}(2014)\citenamefont
  {Nadj-Perge}, \citenamefont {Drozdov}, \citenamefont {Li}, \citenamefont
  {Chen}, \citenamefont {Jeon}, \citenamefont {Seo}, \citenamefont {MacDonald},
  \citenamefont {Bernevig},\ and\ \citenamefont {Yazdani}}]{exp1}%
  \BibitemOpen
  \bibfield  {author} {\bibinfo {author} {\bibfnamefont {S.}~\bibnamefont
  {Nadj-Perge}}, \bibinfo {author} {\bibfnamefont {I.~K.}\ \bibnamefont
  {Drozdov}}, \bibinfo {author} {\bibfnamefont {J.}~\bibnamefont {Li}},
  \bibinfo {author} {\bibfnamefont {H.}~\bibnamefont {Chen}}, \bibinfo {author}
  {\bibfnamefont {S.}~\bibnamefont {Jeon}}, \bibinfo {author} {\bibfnamefont
  {J.}~\bibnamefont {Seo}}, \bibinfo {author} {\bibfnamefont {A.~H.}\
  \bibnamefont {MacDonald}}, \bibinfo {author} {\bibfnamefont {B.~A.}\
  \bibnamefont {Bernevig}}, \ and\ \bibinfo {author} {\bibfnamefont
  {A.}~\bibnamefont {Yazdani}},\ }\href {\doibase 10.1126/science.1259327}
  {\bibfield  {journal} {\bibinfo  {journal} {Science}\ }\textbf {\bibinfo
  {volume} {346}},\ \bibinfo {pages} {602} (\bibinfo {year}
  {2014})}\BibitemShut {NoStop}%
\bibitem [{\citenamefont {Mourik}\ \emph {et~al.}(2012)\citenamefont {Mourik},
  \citenamefont {Zuo}, \citenamefont {Frolov}, \citenamefont {Plissard},
  \citenamefont {Bakkers},\ and\ \citenamefont {Kouwenhoven}}]{exp2}%
  \BibitemOpen
  \bibfield  {author} {\bibinfo {author} {\bibfnamefont {V.}~\bibnamefont
  {Mourik}}, \bibinfo {author} {\bibfnamefont {K.}~\bibnamefont {Zuo}},
  \bibinfo {author} {\bibfnamefont {S.~M.}\ \bibnamefont {Frolov}}, \bibinfo
  {author} {\bibfnamefont {S.~R.}\ \bibnamefont {Plissard}}, \bibinfo {author}
  {\bibfnamefont {E.~P. A.~M.}\ \bibnamefont {Bakkers}}, \ and\ \bibinfo
  {author} {\bibfnamefont {L.~P.}\ \bibnamefont {Kouwenhoven}},\ }\href
  {\doibase 10.1126/science.1222360} {\bibfield  {journal} {\bibinfo  {journal}
  {Science}\ }\textbf {\bibinfo {volume} {336}},\ \bibinfo {pages} {1003}
  (\bibinfo {year} {2012})}\BibitemShut {NoStop}%
\bibitem [{\citenamefont {Sarma}\ \emph {et~al.}(2015)\citenamefont {Sarma},
  \citenamefont {Freedman},\ and\ \citenamefont {Nayak}}]{qtc1}%
  \BibitemOpen
  \bibfield  {author} {\bibinfo {author} {\bibfnamefont {S.~D.}\ \bibnamefont
  {Sarma}}, \bibinfo {author} {\bibfnamefont {M.}~\bibnamefont {Freedman}}, \
  and\ \bibinfo {author} {\bibfnamefont {C.}~\bibnamefont {Nayak}},\ }\href
  {\doibase 10.1038/npjqi.2015.1} {\bibfield  {journal} {\bibinfo  {journal}
  {Npj Quantum Information}\ }\textbf {\bibinfo {volume} {1}},\ \bibinfo
  {pages} {15001} (\bibinfo {year} {2015})}\BibitemShut {NoStop}%
\bibitem [{\citenamefont {Mong}\ \emph {et~al.}(2014)\citenamefont {Mong},
  \citenamefont {Clarke}, \citenamefont {Alicea}, \citenamefont {Lindner},
  \citenamefont {Fendley}, \citenamefont {Nayak}, \citenamefont {Oreg},
  \citenamefont {Stern}, \citenamefont {Berg}, \citenamefont {Shtengel},\ and\
  \citenamefont {Fisher}}]{qtc2}%
  \BibitemOpen
  \bibfield  {author} {\bibinfo {author} {\bibfnamefont {R.~S.~K.}\
  \bibnamefont {Mong}}, \bibinfo {author} {\bibfnamefont {D.~J.}\ \bibnamefont
  {Clarke}}, \bibinfo {author} {\bibfnamefont {J.}~\bibnamefont {Alicea}},
  \bibinfo {author} {\bibfnamefont {N.~H.}\ \bibnamefont {Lindner}}, \bibinfo
  {author} {\bibfnamefont {P.}~\bibnamefont {Fendley}}, \bibinfo {author}
  {\bibfnamefont {C.}~\bibnamefont {Nayak}}, \bibinfo {author} {\bibfnamefont
  {Y.}~\bibnamefont {Oreg}}, \bibinfo {author} {\bibfnamefont {A.}~\bibnamefont
  {Stern}}, \bibinfo {author} {\bibfnamefont {E.}~\bibnamefont {Berg}},
  \bibinfo {author} {\bibfnamefont {K.}~\bibnamefont {Shtengel}}, \ and\
  \bibinfo {author} {\bibfnamefont {M.~P.~A.}\ \bibnamefont {Fisher}},\ }\href
  {\doibase 10.1103/PhysRevX.4.011036} {\bibfield  {journal} {\bibinfo
  {journal} {Phys. Rev. X}\ }\textbf {\bibinfo {volume} {4}},\ \bibinfo {pages}
  {011036} (\bibinfo {year} {2014})}\BibitemShut {NoStop}%
\bibitem [{\citenamefont {Kraus}\ \emph {et~al.}(2013)\citenamefont {Kraus},
  \citenamefont {Zoller},\ and\ \citenamefont {Baranov}}]{qtc3}%
  \BibitemOpen
  \bibfield  {author} {\bibinfo {author} {\bibfnamefont {C.~V.}\ \bibnamefont
  {Kraus}}, \bibinfo {author} {\bibfnamefont {P.}~\bibnamefont {Zoller}}, \
  and\ \bibinfo {author} {\bibfnamefont {M.~A.}\ \bibnamefont {Baranov}},\
  }\href {\doibase 10.1103/PhysRevLett.111.203001} {\bibfield  {journal}
  {\bibinfo  {journal} {Phys. Rev. Lett.}\ }\textbf {\bibinfo {volume} {111}},\
  \bibinfo {pages} {203001} (\bibinfo {year} {2013})}\BibitemShut {NoStop}%
\bibitem [{\citenamefont {Alicea}(2012)}]{qtc4}%
  \BibitemOpen
  \bibfield  {author} {\bibinfo {author} {\bibfnamefont {J.}~\bibnamefont
  {Alicea}},\ }\href {\doibase 10.1088/0034-4885/75/7/076501} {\bibfield
  {journal} {\bibinfo  {journal} {Reports on Progress in Physics}\ }\textbf
  {\bibinfo {volume} {75}},\ \bibinfo {pages} {076501} (\bibinfo {year}
  {2012})}\BibitemShut {NoStop}%
\bibitem [{\citenamefont {Kitaev}(2003)}]{qtc5}%
  \BibitemOpen
  \bibfield  {author} {\bibinfo {author} {\bibfnamefont {A.}~\bibnamefont
  {Kitaev}},\ }\href {\doibase https://doi.org/10.1016/S0003-4916(02)00018-0}
  {\bibfield  {journal} {\bibinfo  {journal} {Annals of Physics}\ }\textbf
  {\bibinfo {volume} {303}},\ \bibinfo {pages} {2 } (\bibinfo {year}
  {2003})}\BibitemShut {NoStop}%
\bibitem [{\citenamefont {Viyuela}\ \emph {et~al.}(2016)\citenamefont
  {Viyuela}, \citenamefont {Vodola}, \citenamefont {Pupillo},\ and\
  \citenamefont {Martin-Delgado}}]{long_range}%
  \BibitemOpen
  \bibfield  {author} {\bibinfo {author} {\bibfnamefont {O.}~\bibnamefont
  {Viyuela}}, \bibinfo {author} {\bibfnamefont {D.}~\bibnamefont {Vodola}},
  \bibinfo {author} {\bibfnamefont {G.}~\bibnamefont {Pupillo}}, \ and\
  \bibinfo {author} {\bibfnamefont {M.~A.}\ \bibnamefont {Martin-Delgado}},\
  }\href {\doibase 10.1103/PhysRevB.94.125121} {\bibfield  {journal} {\bibinfo
  {journal} {Phys. Rev. B}\ }\textbf {\bibinfo {volume} {94}},\ \bibinfo
  {pages} {125121} (\bibinfo {year} {2016})}\BibitemShut {NoStop}%
\bibitem [{\citenamefont {Choi}\ \emph {et~al.}(2017)\citenamefont {Choi},
  \citenamefont {Rubio-Verdú}, \citenamefont {de~Bruijckere}, \citenamefont
  {Ugeda}, \citenamefont {Lorente},\ and\ \citenamefont {Pascual}}]{exp3}%
  \BibitemOpen
  \bibfield  {author} {\bibinfo {author} {\bibfnamefont {D.-J.}\ \bibnamefont
  {Choi}}, \bibinfo {author} {\bibfnamefont {C.}~\bibnamefont {Rubio-Verdú}},
  \bibinfo {author} {\bibfnamefont {J.}~\bibnamefont {de~Bruijckere}}, \bibinfo
  {author} {\bibfnamefont {M.~M.}\ \bibnamefont {Ugeda}}, \bibinfo {author}
  {\bibfnamefont {N.}~\bibnamefont {Lorente}}, \ and\ \bibinfo {author}
  {\bibfnamefont {J.~I.}\ \bibnamefont {Pascual}},\ }\href {\doibase
  10.1038/ncomms15175} {\bibfield  {journal} {\bibinfo  {journal} {Nature
  Communications}\ }\textbf {\bibinfo {volume} {8}},\ \bibinfo {pages} {15175}
  (\bibinfo {year} {2017})}\BibitemShut {NoStop}%
\bibitem [{\citenamefont {Ménard}\ \emph {et~al.}(2017)\citenamefont
  {Ménard}, \citenamefont {Guissart}, \citenamefont {Brun}, \citenamefont
  {Leriche}, \citenamefont {Trif}, \citenamefont {Debontridder}, \citenamefont
  {Demaille}, \citenamefont {Roditchev}, \citenamefont {Simon},\ and\
  \citenamefont {Cren}}]{exp4}%
  \BibitemOpen
  \bibfield  {author} {\bibinfo {author} {\bibfnamefont {G.~C.}\ \bibnamefont
  {Ménard}}, \bibinfo {author} {\bibfnamefont {S.}~\bibnamefont {Guissart}},
  \bibinfo {author} {\bibfnamefont {C.}~\bibnamefont {Brun}}, \bibinfo {author}
  {\bibfnamefont {R.~T.}\ \bibnamefont {Leriche}}, \bibinfo {author}
  {\bibfnamefont {M.}~\bibnamefont {Trif}}, \bibinfo {author} {\bibfnamefont
  {F.}~\bibnamefont {Debontridder}}, \bibinfo {author} {\bibfnamefont
  {D.}~\bibnamefont {Demaille}}, \bibinfo {author} {\bibfnamefont
  {D.}~\bibnamefont {Roditchev}}, \bibinfo {author} {\bibfnamefont
  {P.}~\bibnamefont {Simon}}, \ and\ \bibinfo {author} {\bibfnamefont
  {T.}~\bibnamefont {Cren}},\ }\href {\doibase 10.1038/s41467-017-02192-x}
  {\bibfield  {journal} {\bibinfo  {journal} {Nature Communications}\ }\textbf
  {\bibinfo {volume} {8}},\ \bibinfo {pages} {2040} (\bibinfo {year}
  {2017})}\BibitemShut {NoStop}%
\bibitem [{\citenamefont {Ménard}\ \emph {et~al.}(2015)\citenamefont
  {Ménard}, \citenamefont {Guissart}, \citenamefont {Brun}, \citenamefont
  {Pons}, \citenamefont {Stolyarov}, \citenamefont {Debontridder},
  \citenamefont {Leclerc}, \citenamefont {Janod}, \citenamefont {Cario},
  \citenamefont {Roditchev}, \citenamefont {Simon},\ and\ \citenamefont
  {Cren}}]{exp5}%
  \BibitemOpen
  \bibfield  {author} {\bibinfo {author} {\bibfnamefont {G.~C.}\ \bibnamefont
  {Ménard}}, \bibinfo {author} {\bibfnamefont {S.}~\bibnamefont {Guissart}},
  \bibinfo {author} {\bibfnamefont {C.}~\bibnamefont {Brun}}, \bibinfo {author}
  {\bibfnamefont {S.}~\bibnamefont {Pons}}, \bibinfo {author} {\bibfnamefont
  {V.~S.}\ \bibnamefont {Stolyarov}}, \bibinfo {author} {\bibfnamefont
  {F.}~\bibnamefont {Debontridder}}, \bibinfo {author} {\bibfnamefont {M.~V.}\
  \bibnamefont {Leclerc}}, \bibinfo {author} {\bibfnamefont {E.}~\bibnamefont
  {Janod}}, \bibinfo {author} {\bibfnamefont {L.}~\bibnamefont {Cario}},
  \bibinfo {author} {\bibfnamefont {D.}~\bibnamefont {Roditchev}}, \bibinfo
  {author} {\bibfnamefont {P.}~\bibnamefont {Simon}}, \ and\ \bibinfo {author}
  {\bibfnamefont {T.}~\bibnamefont {Cren}},\ }\href {\doibase
  10.1038/nphys3508} {\bibfield  {journal} {\bibinfo  {journal} {Nature
  Physics}\ }\textbf {\bibinfo {volume} {11}},\ \bibinfo {pages} {1013}
  (\bibinfo {year} {2015})}\BibitemShut {NoStop}%
\bibitem [{\citenamefont {Neupert}\ \emph {et~al.}(2016)\citenamefont
  {Neupert}, \citenamefont {Yazdani},\ and\ \citenamefont {Bernevig}}]{mag1}%
  \BibitemOpen
  \bibfield  {author} {\bibinfo {author} {\bibfnamefont {T.}~\bibnamefont
  {Neupert}}, \bibinfo {author} {\bibfnamefont {A.}~\bibnamefont {Yazdani}}, \
  and\ \bibinfo {author} {\bibfnamefont {B.~A.}\ \bibnamefont {Bernevig}},\
  }\href {\doibase 10.1103/PhysRevB.93.094508} {\bibfield  {journal} {\bibinfo
  {journal} {Phys. Rev. B}\ }\textbf {\bibinfo {volume} {93}},\ \bibinfo
  {pages} {094508} (\bibinfo {year} {2016})}\BibitemShut {NoStop}%
\bibitem [{\citenamefont {Pientka}\ \emph {et~al.}(2014)\citenamefont
  {Pientka}, \citenamefont {Glazman},\ and\ \citenamefont {von Oppen}}]{mag2}%
  \BibitemOpen
  \bibfield  {author} {\bibinfo {author} {\bibfnamefont {F.}~\bibnamefont
  {Pientka}}, \bibinfo {author} {\bibfnamefont {L.~I.}\ \bibnamefont
  {Glazman}}, \ and\ \bibinfo {author} {\bibfnamefont {F.}~\bibnamefont {von
  Oppen}},\ }\href {\doibase 10.1103/PhysRevB.89.180505} {\bibfield  {journal}
  {\bibinfo  {journal} {Phys. Rev. B}\ }\textbf {\bibinfo {volume} {89}},\
  \bibinfo {pages} {180505(R)} (\bibinfo {year} {2014})}\BibitemShut {NoStop}%
\bibitem [{\citenamefont {Klinovaja}\ \emph {et~al.}(2013)\citenamefont
  {Klinovaja}, \citenamefont {Stano}, \citenamefont {Yazdani},\ and\
  \citenamefont {Loss}}]{mag3}%
  \BibitemOpen
  \bibfield  {author} {\bibinfo {author} {\bibfnamefont {J.}~\bibnamefont
  {Klinovaja}}, \bibinfo {author} {\bibfnamefont {P.}~\bibnamefont {Stano}},
  \bibinfo {author} {\bibfnamefont {A.}~\bibnamefont {Yazdani}}, \ and\
  \bibinfo {author} {\bibfnamefont {D.}~\bibnamefont {Loss}},\ }\href {\doibase
  10.1103/PhysRevLett.111.186805} {\bibfield  {journal} {\bibinfo  {journal}
  {Phys. Rev. Lett.}\ }\textbf {\bibinfo {volume} {111}},\ \bibinfo {pages}
  {186805} (\bibinfo {year} {2013})}\BibitemShut {NoStop}%
\bibitem [{\citenamefont {Nadj-Perge}\ \emph {et~al.}(2013)\citenamefont
  {Nadj-Perge}, \citenamefont {Drozdov}, \citenamefont {Bernevig},\ and\
  \citenamefont {Yazdani}}]{mag4}%
  \BibitemOpen
  \bibfield  {author} {\bibinfo {author} {\bibfnamefont {S.}~\bibnamefont
  {Nadj-Perge}}, \bibinfo {author} {\bibfnamefont {I.~K.}\ \bibnamefont
  {Drozdov}}, \bibinfo {author} {\bibfnamefont {B.~A.}\ \bibnamefont
  {Bernevig}}, \ and\ \bibinfo {author} {\bibfnamefont {A.}~\bibnamefont
  {Yazdani}},\ }\href {\doibase 10.1103/PhysRevB.88.020407} {\bibfield
  {journal} {\bibinfo  {journal} {Phys. Rev. B}\ }\textbf {\bibinfo {volume}
  {88}},\ \bibinfo {pages} {020407(R)} (\bibinfo {year} {2013})}\BibitemShut
  {NoStop}%
\bibitem [{\citenamefont {Pientka}\ \emph {et~al.}(2013)\citenamefont
  {Pientka}, \citenamefont {Glazman},\ and\ \citenamefont {von Oppen}}]{mag5}%
  \BibitemOpen
  \bibfield  {author} {\bibinfo {author} {\bibfnamefont {F.}~\bibnamefont
  {Pientka}}, \bibinfo {author} {\bibfnamefont {L.~I.}\ \bibnamefont
  {Glazman}}, \ and\ \bibinfo {author} {\bibfnamefont {F.}~\bibnamefont {von
  Oppen}},\ }\href {\doibase 10.1103/PhysRevB.88.155420} {\bibfield  {journal}
  {\bibinfo  {journal} {Phys. Rev. B}\ }\textbf {\bibinfo {volume} {88}},\
  \bibinfo {pages} {155420} (\bibinfo {year} {2013})}\BibitemShut {NoStop}%
\bibitem [{\citenamefont {Kaladzhyan}\ \emph {et~al.}(2016)\citenamefont
  {Kaladzhyan}, \citenamefont {Bena},\ and\ \citenamefont {Simon}}]{mag6}%
  \BibitemOpen
  \bibfield  {author} {\bibinfo {author} {\bibfnamefont {V.}~\bibnamefont
  {Kaladzhyan}}, \bibinfo {author} {\bibfnamefont {C.}~\bibnamefont {Bena}}, \
  and\ \bibinfo {author} {\bibfnamefont {P.}~\bibnamefont {Simon}},\ }\href
  {\doibase 10.1088/0953-8984/28/48/485701} {\bibfield  {journal} {\bibinfo
  {journal} {Journal of Physics: Condensed Matter}\ }\textbf {\bibinfo {volume}
  {28}},\ \bibinfo {pages} {485701} (\bibinfo {year} {2016})}\BibitemShut
  {NoStop}%
\bibitem [{\citenamefont {Heinrich}\ \emph {et~al.}(2018)\citenamefont
  {Heinrich}, \citenamefont {Pascual},\ and\ \citenamefont {Franke}}]{mag7}%
  \BibitemOpen
  \bibfield  {author} {\bibinfo {author} {\bibfnamefont {B.~W.}\ \bibnamefont
  {Heinrich}}, \bibinfo {author} {\bibfnamefont {J.~I.}\ \bibnamefont
  {Pascual}}, \ and\ \bibinfo {author} {\bibfnamefont {K.~J.}\ \bibnamefont
  {Franke}},\ }\href {\doibase https://doi.org/10.1016/j.progsurf.2018.01.001}
  {\bibfield  {journal} {\bibinfo  {journal} {Progress in Surface Science}\
  }\textbf {\bibinfo {volume} {93}},\ \bibinfo {pages} {1 } (\bibinfo {year}
  {2018})}\BibitemShut {NoStop}%
\bibitem [{\citenamefont {Li}\ and\ \citenamefont {Sun}(2018)}]{coherence}%
  \BibitemOpen
  \bibfield  {author} {\bibinfo {author} {\bibfnamefont {S.-P.}\ \bibnamefont
  {Li}}\ and\ \bibinfo {author} {\bibfnamefont {Z.-H.}\ \bibnamefont {Sun}},\
  }\href {\doibase 10.1103/PhysRevA.98.022317} {\bibfield  {journal} {\bibinfo
  {journal} {Phys. Rev. A}\ }\textbf {\bibinfo {volume} {98}},\ \bibinfo
  {pages} {022317} (\bibinfo {year} {2018})}\BibitemShut {NoStop}%
\bibitem [{\citenamefont {Qiao}\ \emph {et~al.}(2019)\citenamefont {Qiao},
  \citenamefont {Sun}, \citenamefont {Sun}, \citenamefont {Mu}, \citenamefont
  {He},\ and\ \citenamefont {Fan}}]{coherence_add}%
  \BibitemOpen
  \bibfield  {author} {\bibinfo {author} {\bibfnamefont {H.}~\bibnamefont
  {Qiao}}, \bibinfo {author} {\bibfnamefont {Z.-H.}\ \bibnamefont {Sun}},
  \bibinfo {author} {\bibfnamefont {F.-X.}\ \bibnamefont {Sun}}, \bibinfo
  {author} {\bibfnamefont {L.-Z.}\ \bibnamefont {Mu}}, \bibinfo {author}
  {\bibfnamefont {Q.}~\bibnamefont {He}}, \ and\ \bibinfo {author}
  {\bibfnamefont {H.}~\bibnamefont {Fan}},\ }\href {\doibase
  https://doi.org/10.1016/j.aop.2019.167967} {\bibfield  {journal} {\bibinfo
  {journal} {Annals of Physics}\ }\textbf {\bibinfo {volume} {411}},\ \bibinfo
  {pages} {167967} (\bibinfo {year} {2019})}\BibitemShut {NoStop}%
\bibitem [{\citenamefont {Pezz\`e}\ \emph {et~al.}(2017)\citenamefont
  {Pezz\`e}, \citenamefont {Gabbrielli}, \citenamefont {Lepori},\ and\
  \citenamefont {Smerzi}}]{QFI1}%
  \BibitemOpen
  \bibfield  {author} {\bibinfo {author} {\bibfnamefont {L.}~\bibnamefont
  {Pezz\`e}}, \bibinfo {author} {\bibfnamefont {M.}~\bibnamefont {Gabbrielli}},
  \bibinfo {author} {\bibfnamefont {L.}~\bibnamefont {Lepori}}, \ and\ \bibinfo
  {author} {\bibfnamefont {A.}~\bibnamefont {Smerzi}},\ }\href {\doibase
  10.1103/PhysRevLett.119.250401} {\bibfield  {journal} {\bibinfo  {journal}
  {Phys. Rev. Lett.}\ }\textbf {\bibinfo {volume} {119}},\ \bibinfo {pages}
  {250401} (\bibinfo {year} {2017})}\BibitemShut {NoStop}%
\bibitem [{\citenamefont {Zhang}\ \emph
  {et~al.}(2018{\natexlab{a}})\citenamefont {Zhang}, \citenamefont {Zeng},
  \citenamefont {Fan}, \citenamefont {You},\ and\ \citenamefont {Nori}}]{QFI2}%
  \BibitemOpen
  \bibfield  {author} {\bibinfo {author} {\bibfnamefont {Y.-R.}\ \bibnamefont
  {Zhang}}, \bibinfo {author} {\bibfnamefont {Y.}~\bibnamefont {Zeng}},
  \bibinfo {author} {\bibfnamefont {H.}~\bibnamefont {Fan}}, \bibinfo {author}
  {\bibfnamefont {J.~Q.}\ \bibnamefont {You}}, \ and\ \bibinfo {author}
  {\bibfnamefont {F.}~\bibnamefont {Nori}},\ }\href {\doibase
  10.1103/PhysRevLett.120.250501} {\bibfield  {journal} {\bibinfo  {journal}
  {Phys. Rev. Lett.}\ }\textbf {\bibinfo {volume} {120}},\ \bibinfo {pages}
  {250501} (\bibinfo {year} {2018}{\natexlab{a}})}\BibitemShut {NoStop}%
\bibitem [{\citenamefont {Gabbrielli}\ \emph {et~al.}(2018)\citenamefont
  {Gabbrielli}, \citenamefont {Smerzi},\ and\ \citenamefont {Pezzè}}]{QFI3}%
  \BibitemOpen
  \bibfield  {author} {\bibinfo {author} {\bibfnamefont {M.}~\bibnamefont
  {Gabbrielli}}, \bibinfo {author} {\bibfnamefont {A.}~\bibnamefont {Smerzi}},
  \ and\ \bibinfo {author} {\bibfnamefont {L.}~\bibnamefont {Pezzè}},\ }\href
  {\doibase 10.1038/s41598-018-31761-3} {\bibfield  {journal} {\bibinfo
  {journal} {Scientific Reports}\ }\textbf {\bibinfo {volume} {8}},\ \bibinfo
  {pages} {15663} (\bibinfo {year} {2018})}\BibitemShut {NoStop}%
\bibitem [{\citenamefont {Yin}\ \emph {et~al.}(2019)\citenamefont {Yin},
  \citenamefont {Song}, \citenamefont {Zhang},\ and\ \citenamefont
  {Liu}}]{QFI4}%
  \BibitemOpen
  \bibfield  {author} {\bibinfo {author} {\bibfnamefont {S.}~\bibnamefont
  {Yin}}, \bibinfo {author} {\bibfnamefont {J.}~\bibnamefont {Song}}, \bibinfo
  {author} {\bibfnamefont {Y.}~\bibnamefont {Zhang}}, \ and\ \bibinfo {author}
  {\bibfnamefont {S.}~\bibnamefont {Liu}},\ }\href {\doibase
  10.1103/PhysRevB.100.184417} {\bibfield  {journal} {\bibinfo  {journal}
  {Phys. Rev. B}\ }\textbf {\bibinfo {volume} {100}},\ \bibinfo {pages}
  {184417} (\bibinfo {year} {2019})}\BibitemShut {NoStop}%
\bibitem [{\citenamefont {Kitaev}\ and\ \citenamefont {Preskill}(2006)}]{EE1}%
  \BibitemOpen
  \bibfield  {author} {\bibinfo {author} {\bibfnamefont {A.}~\bibnamefont
  {Kitaev}}\ and\ \bibinfo {author} {\bibfnamefont {J.}~\bibnamefont
  {Preskill}},\ }\href {\doibase 10.1103/PhysRevLett.96.110404} {\bibfield
  {journal} {\bibinfo  {journal} {Phys. Rev. Lett.}\ }\textbf {\bibinfo
  {volume} {96}},\ \bibinfo {pages} {110404} (\bibinfo {year}
  {2006})}\BibitemShut {NoStop}%
\bibitem [{\citenamefont {Levin}\ and\ \citenamefont {Wen}(2006)}]{EE2}%
  \BibitemOpen
  \bibfield  {author} {\bibinfo {author} {\bibfnamefont {M.}~\bibnamefont
  {Levin}}\ and\ \bibinfo {author} {\bibfnamefont {X.-G.}\ \bibnamefont
  {Wen}},\ }\href {\doibase 10.1103/PhysRevLett.96.110405} {\bibfield
  {journal} {\bibinfo  {journal} {Phys. Rev. Lett.}\ }\textbf {\bibinfo
  {volume} {96}},\ \bibinfo {pages} {110405} (\bibinfo {year}
  {2006})}\BibitemShut {NoStop}%
\bibitem [{\citenamefont {Vodola}\ \emph {et~al.}(2014)\citenamefont {Vodola},
  \citenamefont {Lepori}, \citenamefont {Ercolessi}, \citenamefont {Gorshkov},\
  and\ \citenamefont {Pupillo}}]{EE3}%
  \BibitemOpen
  \bibfield  {author} {\bibinfo {author} {\bibfnamefont {D.}~\bibnamefont
  {Vodola}}, \bibinfo {author} {\bibfnamefont {L.}~\bibnamefont {Lepori}},
  \bibinfo {author} {\bibfnamefont {E.}~\bibnamefont {Ercolessi}}, \bibinfo
  {author} {\bibfnamefont {A.~V.}\ \bibnamefont {Gorshkov}}, \ and\ \bibinfo
  {author} {\bibfnamefont {G.}~\bibnamefont {Pupillo}},\ }\href {\doibase
  10.1103/PhysRevLett.113.156402} {\bibfield  {journal} {\bibinfo  {journal}
  {Phys. Rev. Lett.}\ }\textbf {\bibinfo {volume} {113}},\ \bibinfo {pages}
  {156402} (\bibinfo {year} {2014})}\BibitemShut {NoStop}%
\bibitem [{\citenamefont {Fromholz}\ \emph {et~al.}(2019)\citenamefont
  {Fromholz}, \citenamefont {Magnifico}, \citenamefont {Vitale}, \citenamefont
  {Mendes-Santos},\ and\ \citenamefont {Dalmonte}}]{EE_add}%
  \BibitemOpen
  \bibfield  {author} {\bibinfo {author} {\bibfnamefont {P.}~\bibnamefont
  {Fromholz}}, \bibinfo {author} {\bibfnamefont {G.}~\bibnamefont {Magnifico}},
  \bibinfo {author} {\bibfnamefont {V.}~\bibnamefont {Vitale}}, \bibinfo
  {author} {\bibfnamefont {T.}~\bibnamefont {Mendes-Santos}}, \ and\ \bibinfo
  {author} {\bibfnamefont {M.}~\bibnamefont {Dalmonte}},\ }\href@noop {}
  {\bibfield  {journal} {\bibinfo  {journal} {ArXiv e-prints}\ } (\bibinfo
  {year} {2019})},\ \Eprint {http://arxiv.org/abs/1909.04035} {arXiv:1909.04035
  [quant-ph]} \BibitemShut {NoStop}%
\bibitem [{\citenamefont {Li}\ and\ \citenamefont {Haldane}(2008)}]{ES1}%
  \BibitemOpen
  \bibfield  {author} {\bibinfo {author} {\bibfnamefont {H.}~\bibnamefont
  {Li}}\ and\ \bibinfo {author} {\bibfnamefont {F.~D.~M.}\ \bibnamefont
  {Haldane}},\ }\href {\doibase 10.1103/PhysRevLett.101.010504} {\bibfield
  {journal} {\bibinfo  {journal} {Phys. Rev. Lett.}\ }\textbf {\bibinfo
  {volume} {101}},\ \bibinfo {pages} {010504} (\bibinfo {year}
  {2008})}\BibitemShut {NoStop}%
\bibitem [{\citenamefont {Pollmann}\ \emph {et~al.}(2010)\citenamefont
  {Pollmann}, \citenamefont {Turner}, \citenamefont {Berg},\ and\ \citenamefont
  {Oshikawa}}]{ES2}%
  \BibitemOpen
  \bibfield  {author} {\bibinfo {author} {\bibfnamefont {F.}~\bibnamefont
  {Pollmann}}, \bibinfo {author} {\bibfnamefont {A.~M.}\ \bibnamefont
  {Turner}}, \bibinfo {author} {\bibfnamefont {E.}~\bibnamefont {Berg}}, \ and\
  \bibinfo {author} {\bibfnamefont {M.}~\bibnamefont {Oshikawa}},\ }\href
  {\doibase 10.1103/PhysRevB.81.064439} {\bibfield  {journal} {\bibinfo
  {journal} {Phys. Rev. B}\ }\textbf {\bibinfo {volume} {81}},\ \bibinfo
  {pages} {064439} (\bibinfo {year} {2010})}\BibitemShut {NoStop}%
\bibitem [{\citenamefont {Fidkowski}(2010)}]{ES3}%
  \BibitemOpen
  \bibfield  {author} {\bibinfo {author} {\bibfnamefont {L.}~\bibnamefont
  {Fidkowski}},\ }\href {\doibase 10.1103/PhysRevLett.104.130502} {\bibfield
  {journal} {\bibinfo  {journal} {Phys. Rev. Lett.}\ }\textbf {\bibinfo
  {volume} {104}},\ \bibinfo {pages} {130502} (\bibinfo {year}
  {2010})}\BibitemShut {NoStop}%
\bibitem [{\citenamefont {Gross}\ and\ \citenamefont {Bloch}(2017)}]{ua_sum}%
  \BibitemOpen
  \bibfield  {author} {\bibinfo {author} {\bibfnamefont {C.}~\bibnamefont
  {Gross}}\ and\ \bibinfo {author} {\bibfnamefont {I.}~\bibnamefont {Bloch}},\
  }\href {\doibase 10.1126/science.aal3837} {\bibfield  {journal} {\bibinfo
  {journal} {Science}\ }\textbf {\bibinfo {volume} {357}},\ \bibinfo {pages}
  {995} (\bibinfo {year} {2017})}\BibitemShut {NoStop}%
\bibitem [{\citenamefont {Fläschner}\ \emph {et~al.}(2018)\citenamefont
  {Fläschner}, \citenamefont {Vogel}, \citenamefont {Tarnowski}, \citenamefont
  {Rem}, \citenamefont {Lühmann}, \citenamefont {Heyl}, \citenamefont
  {Budich}, \citenamefont {Mathey}, \citenamefont {Sengstock},\ and\
  \citenamefont {Weitenberg}}]{ua1}%
  \BibitemOpen
  \bibfield  {author} {\bibinfo {author} {\bibfnamefont {N.}~\bibnamefont
  {Fläschner}}, \bibinfo {author} {\bibfnamefont {D.}~\bibnamefont {Vogel}},
  \bibinfo {author} {\bibfnamefont {M.}~\bibnamefont {Tarnowski}}, \bibinfo
  {author} {\bibfnamefont {B.~S.}\ \bibnamefont {Rem}}, \bibinfo {author}
  {\bibfnamefont {D.-S.}\ \bibnamefont {Lühmann}}, \bibinfo {author}
  {\bibfnamefont {M.}~\bibnamefont {Heyl}}, \bibinfo {author} {\bibfnamefont
  {J.~C.}\ \bibnamefont {Budich}}, \bibinfo {author} {\bibfnamefont
  {L.}~\bibnamefont {Mathey}}, \bibinfo {author} {\bibfnamefont
  {K.}~\bibnamefont {Sengstock}}, \ and\ \bibinfo {author} {\bibfnamefont
  {C.}~\bibnamefont {Weitenberg}},\ }\href {\doibase 10.1038/s41567-017-0013-8}
  {\bibfield  {journal} {\bibinfo  {journal} {Nature Physics}\ }\textbf
  {\bibinfo {volume} {14}},\ \bibinfo {pages} {265} (\bibinfo {year}
  {2018})}\BibitemShut {NoStop}%
\bibitem [{\citenamefont {Sun}\ \emph {et~al.}(2018)\citenamefont {Sun},
  \citenamefont {Yi}, \citenamefont {Wang}, \citenamefont {Zhang},
  \citenamefont {Sanders}, \citenamefont {Xu}, \citenamefont {Wang},
  \citenamefont {Schmiedmayer}, \citenamefont {Deng}, \citenamefont {Liu},
  \citenamefont {Chen},\ and\ \citenamefont {Pan}}]{ua2}%
  \BibitemOpen
  \bibfield  {author} {\bibinfo {author} {\bibfnamefont {W.}~\bibnamefont
  {Sun}}, \bibinfo {author} {\bibfnamefont {C.-R.}\ \bibnamefont {Yi}},
  \bibinfo {author} {\bibfnamefont {B.-Z.}\ \bibnamefont {Wang}}, \bibinfo
  {author} {\bibfnamefont {W.-W.}\ \bibnamefont {Zhang}}, \bibinfo {author}
  {\bibfnamefont {B.~C.}\ \bibnamefont {Sanders}}, \bibinfo {author}
  {\bibfnamefont {X.-T.}\ \bibnamefont {Xu}}, \bibinfo {author} {\bibfnamefont
  {Z.-Y.}\ \bibnamefont {Wang}}, \bibinfo {author} {\bibfnamefont
  {J.}~\bibnamefont {Schmiedmayer}}, \bibinfo {author} {\bibfnamefont
  {Y.}~\bibnamefont {Deng}}, \bibinfo {author} {\bibfnamefont {X.-J.}\
  \bibnamefont {Liu}}, \bibinfo {author} {\bibfnamefont {S.}~\bibnamefont
  {Chen}}, \ and\ \bibinfo {author} {\bibfnamefont {J.-W.}\ \bibnamefont
  {Pan}},\ }\href {\doibase 10.1103/PhysRevLett.121.250403} {\bibfield
  {journal} {\bibinfo  {journal} {Phys. Rev. Lett.}\ }\textbf {\bibinfo
  {volume} {121}},\ \bibinfo {pages} {250403} (\bibinfo {year}
  {2018})}\BibitemShut {NoStop}%
\bibitem [{\citenamefont {Blatt}\ and\ \citenamefont {Roos}(2012)}]{ti_sum}%
  \BibitemOpen
  \bibfield  {author} {\bibinfo {author} {\bibfnamefont {R.}~\bibnamefont
  {Blatt}}\ and\ \bibinfo {author} {\bibfnamefont {C.~F.}\ \bibnamefont
  {Roos}},\ }\href {\doibase 10.1038/nphys2252} {\bibfield  {journal} {\bibinfo
   {journal} {Nature Physics}\ }\textbf {\bibinfo {volume} {8}},\ \bibinfo
  {pages} {277} (\bibinfo {year} {2012})}\BibitemShut {NoStop}%
\bibitem [{\citenamefont {Neill}\ \emph {et~al.}(2017)\citenamefont {Neill},
  \citenamefont {Roushan}, \citenamefont {Fang}, \citenamefont {Chen},
  \citenamefont {Kolodrubetz}, \citenamefont {Chen}, \citenamefont {Megrant},
  \citenamefont {Barends}, \citenamefont {Campbell}, \citenamefont {Chiaro},
  \citenamefont {Dunsworth}, \citenamefont {Jeffrey}, \citenamefont {Kelly},
  \citenamefont {Mutus}, \citenamefont {O’Malley}, \citenamefont {Quintana},
  \citenamefont {Sank}, \citenamefont {Vainsencher}, \citenamefont {Wenner},
  \citenamefont {White}, \citenamefont {Polkovnikov},\ and\ \citenamefont
  {Martinis}}]{sq1}%
  \BibitemOpen
  \bibfield  {author} {\bibinfo {author} {\bibfnamefont {C.}~\bibnamefont
  {Neill}}, \bibinfo {author} {\bibfnamefont {P.}~\bibnamefont {Roushan}},
  \bibinfo {author} {\bibfnamefont {M.}~\bibnamefont {Fang}}, \bibinfo {author}
  {\bibfnamefont {Y.}~\bibnamefont {Chen}}, \bibinfo {author} {\bibfnamefont
  {M.}~\bibnamefont {Kolodrubetz}}, \bibinfo {author} {\bibfnamefont
  {Z.}~\bibnamefont {Chen}}, \bibinfo {author} {\bibfnamefont {A.}~\bibnamefont
  {Megrant}}, \bibinfo {author} {\bibfnamefont {R.}~\bibnamefont {Barends}},
  \bibinfo {author} {\bibfnamefont {B.}~\bibnamefont {Campbell}}, \bibinfo
  {author} {\bibfnamefont {B.}~\bibnamefont {Chiaro}}, \bibinfo {author}
  {\bibfnamefont {A.}~\bibnamefont {Dunsworth}}, \bibinfo {author}
  {\bibfnamefont {E.}~\bibnamefont {Jeffrey}}, \bibinfo {author} {\bibfnamefont
  {J.}~\bibnamefont {Kelly}}, \bibinfo {author} {\bibfnamefont
  {J.}~\bibnamefont {Mutus}}, \bibinfo {author} {\bibfnamefont {P.~J.~J.}\
  \bibnamefont {O’Malley}}, \bibinfo {author} {\bibfnamefont
  {C.}~\bibnamefont {Quintana}}, \bibinfo {author} {\bibfnamefont
  {D.}~\bibnamefont {Sank}}, \bibinfo {author} {\bibfnamefont {A.}~\bibnamefont
  {Vainsencher}}, \bibinfo {author} {\bibfnamefont {J.}~\bibnamefont {Wenner}},
  \bibinfo {author} {\bibfnamefont {T.~C.}\ \bibnamefont {White}}, \bibinfo
  {author} {\bibfnamefont {A.}~\bibnamefont {Polkovnikov}}, \ and\ \bibinfo
  {author} {\bibfnamefont {J.~M.}\ \bibnamefont {Martinis}},\ }\href {\doibase
  10.1038/nphys3830} {\bibfield  {journal} {\bibinfo  {journal} {Nature
  Physics}\ }\textbf {\bibinfo {volume} {12}},\ \bibinfo {pages} {1037}
  (\bibinfo {year} {2017})}\BibitemShut {NoStop}%
\bibitem [{\citenamefont {Yan}\ \emph {et~al.}(2019)\citenamefont {Yan},
  \citenamefont {Zhang}, \citenamefont {Gong}, \citenamefont {Wu},
  \citenamefont {Zheng}, \citenamefont {Li}, \citenamefont {Wang},
  \citenamefont {Liang}, \citenamefont {Lin}, \citenamefont {Xu}, \citenamefont
  {Guo}, \citenamefont {Sun}, \citenamefont {Peng}, \citenamefont {Xia},
  \citenamefont {Deng}, \citenamefont {Rong}, \citenamefont {You},
  \citenamefont {Nori}, \citenamefont {Fan}, \citenamefont {Zhu},\ and\
  \citenamefont {Pan}}]{sq2}%
  \BibitemOpen
  \bibfield  {author} {\bibinfo {author} {\bibfnamefont {Z.}~\bibnamefont
  {Yan}}, \bibinfo {author} {\bibfnamefont {Y.-R.}\ \bibnamefont {Zhang}},
  \bibinfo {author} {\bibfnamefont {M.}~\bibnamefont {Gong}}, \bibinfo {author}
  {\bibfnamefont {Y.}~\bibnamefont {Wu}}, \bibinfo {author} {\bibfnamefont
  {Y.}~\bibnamefont {Zheng}}, \bibinfo {author} {\bibfnamefont
  {S.}~\bibnamefont {Li}}, \bibinfo {author} {\bibfnamefont {C.}~\bibnamefont
  {Wang}}, \bibinfo {author} {\bibfnamefont {F.}~\bibnamefont {Liang}},
  \bibinfo {author} {\bibfnamefont {J.}~\bibnamefont {Lin}}, \bibinfo {author}
  {\bibfnamefont {Y.}~\bibnamefont {Xu}}, \bibinfo {author} {\bibfnamefont
  {C.}~\bibnamefont {Guo}}, \bibinfo {author} {\bibfnamefont {L.}~\bibnamefont
  {Sun}}, \bibinfo {author} {\bibfnamefont {C.-Z.}\ \bibnamefont {Peng}},
  \bibinfo {author} {\bibfnamefont {K.}~\bibnamefont {Xia}}, \bibinfo {author}
  {\bibfnamefont {H.}~\bibnamefont {Deng}}, \bibinfo {author} {\bibfnamefont
  {H.}~\bibnamefont {Rong}}, \bibinfo {author} {\bibfnamefont {J.~Q.}\
  \bibnamefont {You}}, \bibinfo {author} {\bibfnamefont {F.}~\bibnamefont
  {Nori}}, \bibinfo {author} {\bibfnamefont {H.}~\bibnamefont {Fan}}, \bibinfo
  {author} {\bibfnamefont {X.}~\bibnamefont {Zhu}}, \ and\ \bibinfo {author}
  {\bibfnamefont {J.-W.}\ \bibnamefont {Pan}},\ }\href {\doibase
  10.1126/science.aaw1611} {\bibfield  {journal} {\bibinfo  {journal}
  {Science}\ }\textbf {\bibinfo {volume} {364}},\ \bibinfo {pages} {753}
  (\bibinfo {year} {2019})}\BibitemShut {NoStop}%
\bibitem [{\citenamefont {Wilson}\ \emph {et~al.}(2016)\citenamefont {Wilson},
  \citenamefont {Song},\ and\ \citenamefont {Refael}}]{noneq1}%
  \BibitemOpen
  \bibfield  {author} {\bibinfo {author} {\bibfnamefont {J.~H.}\ \bibnamefont
  {Wilson}}, \bibinfo {author} {\bibfnamefont {J.~C.~W.}\ \bibnamefont {Song}},
  \ and\ \bibinfo {author} {\bibfnamefont {G.}~\bibnamefont {Refael}},\ }\href
  {\doibase 10.1103/PhysRevLett.117.235302} {\bibfield  {journal} {\bibinfo
  {journal} {Phys. Rev. Lett.}\ }\textbf {\bibinfo {volume} {117}},\ \bibinfo
  {pages} {235302} (\bibinfo {year} {2016})}\BibitemShut {NoStop}%
\bibitem [{\citenamefont {Wang}\ \emph {et~al.}(2017)\citenamefont {Wang},
  \citenamefont {Zhang}, \citenamefont {Chen}, \citenamefont {Yu},\ and\
  \citenamefont {Zhai}}]{noneq2}%
  \BibitemOpen
  \bibfield  {author} {\bibinfo {author} {\bibfnamefont {C.}~\bibnamefont
  {Wang}}, \bibinfo {author} {\bibfnamefont {P.}~\bibnamefont {Zhang}},
  \bibinfo {author} {\bibfnamefont {X.}~\bibnamefont {Chen}}, \bibinfo {author}
  {\bibfnamefont {J.}~\bibnamefont {Yu}}, \ and\ \bibinfo {author}
  {\bibfnamefont {H.}~\bibnamefont {Zhai}},\ }\href {\doibase
  10.1103/PhysRevLett.118.185701} {\bibfield  {journal} {\bibinfo  {journal}
  {Phys. Rev. Lett.}\ }\textbf {\bibinfo {volume} {118}},\ \bibinfo {pages}
  {185701} (\bibinfo {year} {2017})}\BibitemShut {NoStop}%
\bibitem [{\citenamefont {Maffei}\ \emph {et~al.}(2018)\citenamefont {Maffei},
  \citenamefont {Dauphin}, \citenamefont {Cardano}, \citenamefont
  {Lewenstein},\ and\ \citenamefont {Massignan}}]{noneq3}%
  \BibitemOpen
  \bibfield  {author} {\bibinfo {author} {\bibfnamefont {M.}~\bibnamefont
  {Maffei}}, \bibinfo {author} {\bibfnamefont {A.}~\bibnamefont {Dauphin}},
  \bibinfo {author} {\bibfnamefont {F.}~\bibnamefont {Cardano}}, \bibinfo
  {author} {\bibfnamefont {M.}~\bibnamefont {Lewenstein}}, \ and\ \bibinfo
  {author} {\bibfnamefont {P.}~\bibnamefont {Massignan}},\ }\href {\doibase
  10.1088/1367-2630/aa9d4c} {\bibfield  {journal} {\bibinfo  {journal} {New
  Journal of Physics}\ }\textbf {\bibinfo {volume} {20}},\ \bibinfo {pages}
  {013023} (\bibinfo {year} {2018})}\BibitemShut {NoStop}%
\bibitem [{\citenamefont {Zhang}\ \emph
  {et~al.}(2018{\natexlab{b}})\citenamefont {Zhang}, \citenamefont {Zhang},
  \citenamefont {Niu},\ and\ \citenamefont {Liu}}]{noneq4}%
  \BibitemOpen
  \bibfield  {author} {\bibinfo {author} {\bibfnamefont {L.}~\bibnamefont
  {Zhang}}, \bibinfo {author} {\bibfnamefont {L.}~\bibnamefont {Zhang}},
  \bibinfo {author} {\bibfnamefont {S.}~\bibnamefont {Niu}}, \ and\ \bibinfo
  {author} {\bibfnamefont {X.-J.}\ \bibnamefont {Liu}},\ }\href {\doibase
  https://doi.org/10.1016/j.scib.2018.09.018} {\bibfield  {journal} {\bibinfo
  {journal} {Science Bulletin}\ }\textbf {\bibinfo {volume} {63}},\ \bibinfo
  {pages} {1385 } (\bibinfo {year} {2018}{\natexlab{b}})}\BibitemShut {NoStop}%
\bibitem [{\citenamefont {Potter}\ \emph {et~al.}(2016)\citenamefont {Potter},
  \citenamefont {Morimoto},\ and\ \citenamefont {Vishwanath}}]{noneq5}%
  \BibitemOpen
  \bibfield  {author} {\bibinfo {author} {\bibfnamefont {A.~C.}\ \bibnamefont
  {Potter}}, \bibinfo {author} {\bibfnamefont {T.}~\bibnamefont {Morimoto}}, \
  and\ \bibinfo {author} {\bibfnamefont {A.}~\bibnamefont {Vishwanath}},\
  }\href {\doibase 10.1103/PhysRevX.6.041001} {\bibfield  {journal} {\bibinfo
  {journal} {Phys. Rev. X}\ }\textbf {\bibinfo {volume} {6}},\ \bibinfo {pages}
  {041001} (\bibinfo {year} {2016})}\BibitemShut {NoStop}%
\bibitem [{\citenamefont {Su}\ \emph {et~al.}(1979)\citenamefont {Su},
  \citenamefont {Schrieffer},\ and\ \citenamefont {Heeger}}]{SSH}%
  \BibitemOpen
  \bibfield  {author} {\bibinfo {author} {\bibfnamefont {W.~P.}\ \bibnamefont
  {Su}}, \bibinfo {author} {\bibfnamefont {J.~R.}\ \bibnamefont {Schrieffer}},
  \ and\ \bibinfo {author} {\bibfnamefont {A.~J.}\ \bibnamefont {Heeger}},\
  }\href {\doibase 10.1103/PhysRevLett.42.1698} {\bibfield  {journal} {\bibinfo
   {journal} {Phys. Rev. Lett.}\ }\textbf {\bibinfo {volume} {42}},\ \bibinfo
  {pages} {1698} (\bibinfo {year} {1979})}\BibitemShut {NoStop}%
\bibitem [{\citenamefont {Kitaev}(2001)}]{Kitaevchain}%
  \BibitemOpen
  \bibfield  {author} {\bibinfo {author} {\bibfnamefont {A.~Y.}\ \bibnamefont
  {Kitaev}},\ }\href {\doibase 10.1070/1063-7869/44/10s/s29} {\bibfield
  {journal} {\bibinfo  {journal} {Physics-Uspekhi}\ }\textbf {\bibinfo {volume}
  {44}},\ \bibinfo {pages} {131} (\bibinfo {year} {2001})}\BibitemShut
  {NoStop}%
\bibitem [{\citenamefont {Gong}\ and\ \citenamefont {Ueda}(2018)}]{ESnoneq1}%
  \BibitemOpen
  \bibfield  {author} {\bibinfo {author} {\bibfnamefont {Z.}~\bibnamefont
  {Gong}}\ and\ \bibinfo {author} {\bibfnamefont {M.}~\bibnamefont {Ueda}},\
  }\href {\doibase 10.1103/PhysRevLett.121.250601} {\bibfield  {journal}
  {\bibinfo  {journal} {Phys. Rev. Lett.}\ }\textbf {\bibinfo {volume} {121}},\
  \bibinfo {pages} {250601} (\bibinfo {year} {2018})}\BibitemShut {NoStop}%
\bibitem [{\citenamefont {Chang}(2018)}]{ESnoneq2}%
  \BibitemOpen
  \bibfield  {author} {\bibinfo {author} {\bibfnamefont {P.-Y.}\ \bibnamefont
  {Chang}},\ }\href {\doibase 10.1103/PhysRevB.97.224304} {\bibfield  {journal}
  {\bibinfo  {journal} {Phys. Rev. B}\ }\textbf {\bibinfo {volume} {97}},\
  \bibinfo {pages} {224304} (\bibinfo {year} {2018})}\BibitemShut {NoStop}%
\bibitem [{\citenamefont {Lu}\ and\ \citenamefont {Yu}(2019)}]{ESnoneq3}%
  \BibitemOpen
  \bibfield  {author} {\bibinfo {author} {\bibfnamefont {S.}~\bibnamefont
  {Lu}}\ and\ \bibinfo {author} {\bibfnamefont {J.}~\bibnamefont {Yu}},\ }\href
  {\doibase 10.1103/PhysRevA.99.033621} {\bibfield  {journal} {\bibinfo
  {journal} {Phys. Rev. A}\ }\textbf {\bibinfo {volume} {99}},\ \bibinfo
  {pages} {033621} (\bibinfo {year} {2019})}\BibitemShut {NoStop}%
\bibitem [{\citenamefont {Xu}\ \emph {et~al.}(2018)\citenamefont {Xu},
  \citenamefont {Chen}, \citenamefont {Zeng}, \citenamefont {Zhang},
  \citenamefont {Song}, \citenamefont {Liu}, \citenamefont {Guo}, \citenamefont
  {Zhang}, \citenamefont {Xu}, \citenamefont {Deng}, \citenamefont {Huang},
  \citenamefont {Wang}, \citenamefont {Zhu}, \citenamefont {Zheng},\ and\
  \citenamefont {Fan}}]{qst1}%
  \BibitemOpen
  \bibfield  {author} {\bibinfo {author} {\bibfnamefont {K.}~\bibnamefont
  {Xu}}, \bibinfo {author} {\bibfnamefont {J.-J.}\ \bibnamefont {Chen}},
  \bibinfo {author} {\bibfnamefont {Y.}~\bibnamefont {Zeng}}, \bibinfo {author}
  {\bibfnamefont {Y.-R.}\ \bibnamefont {Zhang}}, \bibinfo {author}
  {\bibfnamefont {C.}~\bibnamefont {Song}}, \bibinfo {author} {\bibfnamefont
  {W.}~\bibnamefont {Liu}}, \bibinfo {author} {\bibfnamefont {Q.}~\bibnamefont
  {Guo}}, \bibinfo {author} {\bibfnamefont {P.}~\bibnamefont {Zhang}}, \bibinfo
  {author} {\bibfnamefont {D.}~\bibnamefont {Xu}}, \bibinfo {author}
  {\bibfnamefont {H.}~\bibnamefont {Deng}}, \bibinfo {author} {\bibfnamefont
  {K.}~\bibnamefont {Huang}}, \bibinfo {author} {\bibfnamefont
  {H.}~\bibnamefont {Wang}}, \bibinfo {author} {\bibfnamefont {X.}~\bibnamefont
  {Zhu}}, \bibinfo {author} {\bibfnamefont {D.}~\bibnamefont {Zheng}}, \ and\
  \bibinfo {author} {\bibfnamefont {H.}~\bibnamefont {Fan}},\ }\href {\doibase
  10.1103/PhysRevLett.120.050507} {\bibfield  {journal} {\bibinfo  {journal}
  {Phys. Rev. Lett.}\ }\textbf {\bibinfo {volume} {120}},\ \bibinfo {pages}
  {050507} (\bibinfo {year} {2018})}\BibitemShut {NoStop}%
\bibitem [{\citenamefont {Lanyon}\ \emph {et~al.}(2018)\citenamefont {Lanyon},
  \citenamefont {Maier}, \citenamefont {Holzäpfel}, \citenamefont {Baumgratz},
  \citenamefont {Hempel}, \citenamefont {Jurcevic}, \citenamefont {Dhand},
  \citenamefont {Buyskikh}, \citenamefont {Daley}, \citenamefont {Cramer},
  \citenamefont {Plenio}, \citenamefont {Blatt},\ and\ \citenamefont
  {Roos}}]{qst2}%
  \BibitemOpen
  \bibfield  {author} {\bibinfo {author} {\bibfnamefont {B.~P.}\ \bibnamefont
  {Lanyon}}, \bibinfo {author} {\bibfnamefont {C.}~\bibnamefont {Maier}},
  \bibinfo {author} {\bibfnamefont {M.}~\bibnamefont {Holzäpfel}}, \bibinfo
  {author} {\bibfnamefont {T.}~\bibnamefont {Baumgratz}}, \bibinfo {author}
  {\bibfnamefont {C.}~\bibnamefont {Hempel}}, \bibinfo {author} {\bibfnamefont
  {P.}~\bibnamefont {Jurcevic}}, \bibinfo {author} {\bibfnamefont
  {I.}~\bibnamefont {Dhand}}, \bibinfo {author} {\bibfnamefont {A.~S.}\
  \bibnamefont {Buyskikh}}, \bibinfo {author} {\bibfnamefont {A.~J.}\
  \bibnamefont {Daley}}, \bibinfo {author} {\bibfnamefont {M.}~\bibnamefont
  {Cramer}}, \bibinfo {author} {\bibfnamefont {M.~B.}\ \bibnamefont {Plenio}},
  \bibinfo {author} {\bibfnamefont {R.}~\bibnamefont {Blatt}}, \ and\ \bibinfo
  {author} {\bibfnamefont {C.~F.}\ \bibnamefont {Roos}},\ }\href {\doibase
  10.1038/nphys4244} {\bibfield  {journal} {\bibinfo  {journal} {Nature
  Physics}\ }\textbf {\bibinfo {volume} {13}},\ \bibinfo {pages} {1158}
  (\bibinfo {year} {2018})}\BibitemShut {NoStop}%
\bibitem [{\citenamefont {Choo}\ \emph {et~al.}(2018)\citenamefont {Choo},
  \citenamefont {von Keyserlingk}, \citenamefont {Regnault},\ and\
  \citenamefont {Neupert}}]{qst3}%
  \BibitemOpen
  \bibfield  {author} {\bibinfo {author} {\bibfnamefont {K.}~\bibnamefont
  {Choo}}, \bibinfo {author} {\bibfnamefont {C.~W.}\ \bibnamefont {von
  Keyserlingk}}, \bibinfo {author} {\bibfnamefont {N.}~\bibnamefont
  {Regnault}}, \ and\ \bibinfo {author} {\bibfnamefont {T.}~\bibnamefont
  {Neupert}},\ }\href {\doibase 10.1103/PhysRevLett.121.086808} {\bibfield
  {journal} {\bibinfo  {journal} {Phys. Rev. Lett.}\ }\textbf {\bibinfo
  {volume} {121}},\ \bibinfo {pages} {086808} (\bibinfo {year}
  {2018})}\BibitemShut {NoStop}%
\bibitem [{\citenamefont {Chiu}\ \emph {et~al.}(2016)\citenamefont {Chiu},
  \citenamefont {Teo}, \citenamefont {Schnyder},\ and\ \citenamefont
  {Ryu}}]{winding}%
  \BibitemOpen
  \bibfield  {author} {\bibinfo {author} {\bibfnamefont {C.-K.}\ \bibnamefont
  {Chiu}}, \bibinfo {author} {\bibfnamefont {J.~C.~Y.}\ \bibnamefont {Teo}},
  \bibinfo {author} {\bibfnamefont {A.~P.}\ \bibnamefont {Schnyder}}, \ and\
  \bibinfo {author} {\bibfnamefont {S.}~\bibnamefont {Ryu}},\ }\href {\doibase
  10.1103/RevModPhys.88.035005} {\bibfield  {journal} {\bibinfo  {journal}
  {Rev. Mod. Phys.}\ }\textbf {\bibinfo {volume} {88}},\ \bibinfo {pages}
  {035005} (\bibinfo {year} {2016})}\BibitemShut {NoStop}%
\bibitem [{\citenamefont {Cheong}\ and\ \citenamefont
  {Henley}(2004)}]{EE_Hamiltonian}%
  \BibitemOpen
  \bibfield  {author} {\bibinfo {author} {\bibfnamefont {S.-A.}\ \bibnamefont
  {Cheong}}\ and\ \bibinfo {author} {\bibfnamefont {C.~L.}\ \bibnamefont
  {Henley}},\ }\href {\doibase 10.1103/PhysRevB.69.075111} {\bibfield
  {journal} {\bibinfo  {journal} {Phys. Rev. B}\ }\textbf {\bibinfo {volume}
  {69}},\ \bibinfo {pages} {075111} (\bibinfo {year} {2004})}\BibitemShut
  {NoStop}%
\bibitem [{\citenamefont {Hughes}\ \emph {et~al.}(2011)\citenamefont {Hughes},
  \citenamefont {Prodan},\ and\ \citenamefont {Bernevig}}]{Es_add1}%
  \BibitemOpen
  \bibfield  {author} {\bibinfo {author} {\bibfnamefont {T.~L.}\ \bibnamefont
  {Hughes}}, \bibinfo {author} {\bibfnamefont {E.}~\bibnamefont {Prodan}}, \
  and\ \bibinfo {author} {\bibfnamefont {B.~A.}\ \bibnamefont {Bernevig}},\
  }\href {\doibase 10.1103/PhysRevB.83.245132} {\bibfield  {journal} {\bibinfo
  {journal} {Phys. Rev. B}\ }\textbf {\bibinfo {volume} {83}},\ \bibinfo
  {pages} {245132} (\bibinfo {year} {2011})}\BibitemShut {NoStop}%
\bibitem [{\citenamefont {Kitaev}(2009)}]{Kitaev_add}%
  \BibitemOpen
  \bibfield  {author} {\bibinfo {author} {\bibfnamefont {A.}~\bibnamefont
  {Kitaev}},\ }\href {\doibase 10.1063/1.3149495} {\bibfield  {journal}
  {\bibinfo  {journal} {AIP Conf. Proc.}\ }\textbf {\bibinfo {volume} {1134}},\
  \bibinfo {pages} {22} (\bibinfo {year} {2009})}\BibitemShut {NoStop}%
\bibitem [{\citenamefont {Emary}\ and\ \citenamefont {Brandes}(2003)}]{Dicke}%
  \BibitemOpen
  \bibfield  {author} {\bibinfo {author} {\bibfnamefont {C.}~\bibnamefont
  {Emary}}\ and\ \bibinfo {author} {\bibfnamefont {T.}~\bibnamefont
  {Brandes}},\ }\href {\doibase 10.1103/PhysRevE.67.066203} {\bibfield
  {journal} {\bibinfo  {journal} {Phys. Rev. E}\ }\textbf {\bibinfo {volume}
  {67}},\ \bibinfo {pages} {066203} (\bibinfo {year} {2003})}\BibitemShut
  {NoStop}%
\bibitem [{\citenamefont {Zhang}\ and\ \citenamefont
  {Song}(2015)}]{high_winding1}%
  \BibitemOpen
  \bibfield  {author} {\bibinfo {author} {\bibfnamefont {G.}~\bibnamefont
  {Zhang}}\ and\ \bibinfo {author} {\bibfnamefont {Z.}~\bibnamefont {Song}},\
  }\href {\doibase 10.1103/PhysRevLett.115.177204} {\bibfield  {journal}
  {\bibinfo  {journal} {Phys. Rev. Lett.}\ }\textbf {\bibinfo {volume} {115}},\
  \bibinfo {pages} {177204} (\bibinfo {year} {2015})}\BibitemShut {NoStop}%
\bibitem [{\citenamefont {Alecce}\ and\ \citenamefont
  {Dell'Anna}(2017)}]{high_winding2}%
  \BibitemOpen
  \bibfield  {author} {\bibinfo {author} {\bibfnamefont {A.}~\bibnamefont
  {Alecce}}\ and\ \bibinfo {author} {\bibfnamefont {L.}~\bibnamefont
  {Dell'Anna}},\ }\href {\doibase 10.1103/PhysRevB.95.195160} {\bibfield
  {journal} {\bibinfo  {journal} {Phys. Rev. B}\ }\textbf {\bibinfo {volume}
  {95}},\ \bibinfo {pages} {195160} (\bibinfo {year} {2017})}\BibitemShut
  {NoStop}%
\bibitem [{\citenamefont {Viyuela}\ \emph {et~al.}(2018)\citenamefont
  {Viyuela}, \citenamefont {Fu},\ and\ \citenamefont
  {Martin-Delgado}}]{two_dim}%
  \BibitemOpen
  \bibfield  {author} {\bibinfo {author} {\bibfnamefont {O.}~\bibnamefont
  {Viyuela}}, \bibinfo {author} {\bibfnamefont {L.}~\bibnamefont {Fu}}, \ and\
  \bibinfo {author} {\bibfnamefont {M.~A.}\ \bibnamefont {Martin-Delgado}},\
  }\href {\doibase 10.1103/PhysRevLett.120.017001} {\bibfield  {journal}
  {\bibinfo  {journal} {Phys. Rev. Lett.}\ }\textbf {\bibinfo {volume} {120}},\
  \bibinfo {pages} {017001} (\bibinfo {year} {2018})}\BibitemShut {NoStop}%
\bibitem [{\citenamefont {Koffel}\ \emph {et~al.}(2012)\citenamefont {Koffel},
  \citenamefont {Lewenstein},\ and\ \citenamefont {Tagliacozzo}}]{CQPT}%
  \BibitemOpen
  \bibfield  {author} {\bibinfo {author} {\bibfnamefont {T.}~\bibnamefont
  {Koffel}}, \bibinfo {author} {\bibfnamefont {M.}~\bibnamefont {Lewenstein}},
  \ and\ \bibinfo {author} {\bibfnamefont {L.}~\bibnamefont {Tagliacozzo}},\
  }\href {\doibase 10.1103/PhysRevLett.109.267203} {\bibfield  {journal}
  {\bibinfo  {journal} {Phys. Rev. Lett.}\ }\textbf {\bibinfo {volume} {109}},\
  \bibinfo {pages} {267203} (\bibinfo {year} {2012})}\BibitemShut {NoStop}%
\bibitem [{\citenamefont {\ifmmode \check{Z}\else
  \v{Z}\fi{}unkovi\ifmmode~\check{c}\else \v{c}\fi{}}\ \emph
  {et~al.}(2018)\citenamefont {\ifmmode \check{Z}\else
  \v{Z}\fi{}unkovi\ifmmode~\check{c}\else \v{c}\fi{}}, \citenamefont {Heyl},
  \citenamefont {Knap},\ and\ \citenamefont {Silva}}]{DQPT}%
  \BibitemOpen
  \bibfield  {author} {\bibinfo {author} {\bibfnamefont {B.}~\bibnamefont
  {\ifmmode \check{Z}\else \v{Z}\fi{}unkovi\ifmmode~\check{c}\else
  \v{c}\fi{}}}, \bibinfo {author} {\bibfnamefont {M.}~\bibnamefont {Heyl}},
  \bibinfo {author} {\bibfnamefont {M.}~\bibnamefont {Knap}}, \ and\ \bibinfo
  {author} {\bibfnamefont {A.}~\bibnamefont {Silva}},\ }\href {\doibase
  10.1103/PhysRevLett.120.130601} {\bibfield  {journal} {\bibinfo  {journal}
  {Phys. Rev. Lett.}\ }\textbf {\bibinfo {volume} {120}},\ \bibinfo {pages}
  {130601} (\bibinfo {year} {2018})}\BibitemShut {NoStop}%
\bibitem [{\citenamefont {Xu}\ \emph {et~al.}(2019)\citenamefont {Xu},
  \citenamefont {Sun}, \citenamefont {Liu}, \citenamefont {Zhang},
  \citenamefont {Li}, \citenamefont {Dong}, \citenamefont {Ren}, \citenamefont
  {Zhang}, \citenamefont {Nori}, \citenamefont {Zheng}, \citenamefont {Fan},\
  and\ \citenamefont {Wang}}]{DQPT_exp}%
  \BibitemOpen
  \bibfield  {author} {\bibinfo {author} {\bibfnamefont {K.}~\bibnamefont
  {Xu}}, \bibinfo {author} {\bibfnamefont {Z.-H.}\ \bibnamefont {Sun}},
  \bibinfo {author} {\bibfnamefont {W.}~\bibnamefont {Liu}}, \bibinfo {author}
  {\bibfnamefont {Y.-R.}\ \bibnamefont {Zhang}}, \bibinfo {author}
  {\bibfnamefont {H.}~\bibnamefont {Li}}, \bibinfo {author} {\bibfnamefont
  {H.}~\bibnamefont {Dong}}, \bibinfo {author} {\bibfnamefont {W.}~\bibnamefont
  {Ren}}, \bibinfo {author} {\bibfnamefont {P.}~\bibnamefont {Zhang}}, \bibinfo
  {author} {\bibfnamefont {F.}~\bibnamefont {Nori}}, \bibinfo {author}
  {\bibfnamefont {D.}~\bibnamefont {Zheng}}, \bibinfo {author} {\bibfnamefont
  {H.}~\bibnamefont {Fan}}, \ and\ \bibinfo {author} {\bibfnamefont
  {H.}~\bibnamefont {Wang}},\ }\href@noop {} {\bibfield  {journal} {\bibinfo
  {journal} {ArXiv e-prints}\ } (\bibinfo {year} {2019})},\ \Eprint
  {http://arxiv.org/abs/1912.05150} {arXiv:1912.05150 [quant-ph]} \BibitemShut
  {NoStop}%
\bibitem [{\citenamefont {Sedlmayr}\ \emph {et~al.}(2018)\citenamefont
  {Sedlmayr}, \citenamefont {Jaeger}, \citenamefont {Maiti},\ and\
  \citenamefont {Sirker}}]{DQPT_add}%
  \BibitemOpen
  \bibfield  {author} {\bibinfo {author} {\bibfnamefont {N.}~\bibnamefont
  {Sedlmayr}}, \bibinfo {author} {\bibfnamefont {P.}~\bibnamefont {Jaeger}},
  \bibinfo {author} {\bibfnamefont {M.}~\bibnamefont {Maiti}}, \ and\ \bibinfo
  {author} {\bibfnamefont {J.}~\bibnamefont {Sirker}},\ }\href {\doibase
  10.1103/PhysRevB.97.064304} {\bibfield  {journal} {\bibinfo  {journal} {Phys.
  Rev. B}\ }\textbf {\bibinfo {volume} {97}},\ \bibinfo {pages} {064304}
  (\bibinfo {year} {2018})}\BibitemShut {NoStop}%
\bibitem [{\citenamefont {Masłowski}\ and\ \citenamefont
  {Sedlmayr}(2019)}]{DQPT_add2}%
  \BibitemOpen
  \bibfield  {author} {\bibinfo {author} {\bibfnamefont {T.}~\bibnamefont
  {Masłowski}}\ and\ \bibinfo {author} {\bibfnamefont {N.}~\bibnamefont
  {Sedlmayr}},\ }\href@noop {} {\bibfield  {journal} {\bibinfo  {journal}
  {ArXiv e-prints}\ } (\bibinfo {year} {2019})},\ \Eprint
  {http://arxiv.org/abs/1911.02831} {arXiv:1911.02831 [quant-ph]} \BibitemShut
  {NoStop}%
\bibitem [{\citenamefont {Van~Regemortel}\ \emph {et~al.}(2016)\citenamefont
  {Van~Regemortel}, \citenamefont {Sels},\ and\ \citenamefont {Wouters}}]{Ek}%
  \BibitemOpen
  \bibfield  {author} {\bibinfo {author} {\bibfnamefont {M.}~\bibnamefont
  {Van~Regemortel}}, \bibinfo {author} {\bibfnamefont {D.}~\bibnamefont
  {Sels}}, \ and\ \bibinfo {author} {\bibfnamefont {M.}~\bibnamefont
  {Wouters}},\ }\href {\doibase 10.1103/PhysRevA.93.032311} {\bibfield
  {journal} {\bibinfo  {journal} {Phys. Rev. A}\ }\textbf {\bibinfo {volume}
  {93}},\ \bibinfo {pages} {032311} (\bibinfo {year} {2016})}\BibitemShut
  {NoStop}%
\end{thebibliography}%

\end{document}